\documentclass[runningheads]{llncs}
\usepackage{amsmath}

\usepackage[T1]{fontenc}
\usepackage{graphicx}

\usepackage{epsfig}
\usepackage{amssymb}
\usepackage{amsfonts}
\usepackage{algorithm}
\usepackage{tabu}
\usepackage{multirow}

\usepackage{caption}
\usepackage{subcaption}
\usepackage{changepage}
\usepackage{algpseudocode}
\usepackage{xcolor}
\usepackage[breaklinks=true]{hyperref}
\usepackage{breakcites}
\begin{document}
\title{A Comprehensive Review of Latent Space Dynamics Identification Algorithms for Intrusive and Non-Intrusive Reduced-Order-Modeling}
\titlerunning{A Comprehensive LaSDI Review for Reduced Order Modelling}
%

\author{Christophe Bonneville\inst{1}\textsuperscript{*}\textsuperscript{\textdagger} \and
Xiaolong He\inst{2} \and April Tran \inst{3} \and Jun Sur R.~Park \inst{4} \and William Fries \inst{5} \and Daniel A.~Messenger \inst{3} \and Siu Wun Cheung \inst{6} \and Yeonjong Shin \inst{7, 8} \and David M.~Bortz \inst{3} \and Debojyoti Ghosh \inst{6} \and Jiun-Shyan Chen \inst{9} \and Jonathan Belof \inst{6} \and Youngsoo Choi \inst{6}}
\authorrunning{Bonneville et al.}
%
\institute{Civil \& Environmental Engineering, Cornell University, Ithaca, NY 14850, United States \and
ANSYS Inc., Livermore, CA 94551, United States \and Applied Mathematics, University of Colorado, Boulder, CO 80309, United States \and
Center for Artificial Intelligence and Natural Sciences, Korea Institute
for Advanced Study, Seoul, 02455, Republic of Korea \and
University of Arizona, Tucson, AZ 85721, United States \and Lawrence Livermore National Laboratory, Livermore, CA 94550, United States \and
Department of Mathematics, North Carolina State University, Raleigh,
NC 27695, United States \and Mathematical Institute for Data Science, Pohang University of Science
and Technology, Pohang, 37673, Republic of Korea \and Structural Engineering, University of California, San Diego, CA 
92093, United States}
\maketitle              
\begin{abstract}
Numerical solvers of partial differential equations (PDEs) have been widely employed for simulating physical systems. However, the computational cost remains a major bottleneck in various scientific and engineering applications, which has motivated the development of reduced-order models (ROMs). Recently, machine-learning-based ROMs have gained significant popularity and are promising for addressing some limitations of traditional ROM methods, especially for advection dominated systems. In this chapter, we focus on a particular framework known as \textit{Latent Space Dynamics Identification} (LaSDI), which transforms the high-fidelity data, governed by a PDE, to simpler and low-dimensional latent-space data, governed by ordinary differential equations (ODEs). These ODEs can be learned and subsequently interpolated to make ROM predictions. Each building block of LaSDI can be easily modulated depending on the application, which makes the LaSDI framework highly flexible. In particular, we present strategies to enforce the laws of thermodynamics into LaSDI models (tLaSDI), enhance robustness in the presence of noise through the weak form (WLaSDI), select high-fidelity training data efficiently through active learning (gLaSDI, GPLaSDI), and quantify the ROM prediction uncertainty through Gaussian processes (GPLaSDI). We demonstrate the performance of different LaSDI approaches on Burgers equation, a non-linear heat conduction problem, and a plasma physics problem, showing that LaSDI algorithms can achieve relative errors of less than a few percent and up to thousands of times speed-ups.

\end{abstract}
\let\thefootnote\relax\footnotetext{\textsuperscript{*}Now at Sandia National Laboratories, Livermore, CA 94550, United States\\
\textsuperscript{\textdagger}Corresponding author: \href{cpb97@cornell.edu}{cpb97@cornell.edu}}
\section{Introduction}

In recent years, there has been an impressive growth in the development of numerical simulation methods to comprehend physical phenomena, resulting in enhanced accuracy and level of sophistication. Alongside this, there have been substantial advancements in computational hardware, making it more powerful and cost-effective. Consequently, numerical simulations have become widely adopted in numerous areas, such as engineering design, digital twins, decision-making processes \cite{alma991043311449403276,JONES202036,Journal,article,sep-simulations-science}, and a range of fields including aerospace, automotive, electronics, physics, and biology. \cite{cummings_mason_morton_mcdaniel_2015,6db924dfeff44d159ab577c1aefed6ef,car1,9043275,Peterson_b1998,rylander,thijssen_2007,russel}. 

In the realms of engineering and physics, computational simulations frequently involve the resolution of partial differential equations (PDEs) using various numerical approaches, such as finite difference/volume/element methods, particle methods, and others. Although these techniques are known for their precision and ability to produce detailed simulations when implemented appropriately, they require substantial computational resources. This becomes particularly true in complex, time-sensitive multiscale problems that deal with intricate physical phenomenons (for instance, turbulent fluid flows, the dynamics of plasma in fusion devices, and astrodynamic flows) on high resolution grids and meshes. As a result, conducting numerous high-fidelity simulations with these advanced solvers can present considerable computational challenges, especially in scenarios involving uncertainty quantification \cite{10754/656260,Smith2013UncertaintyQ,Sternfels_2013}, inverse problems \cite{10754/656260,https://doi.org/10.1002/nme.2746,https://doi.org/10.1002/2016RS005998,Sternfels_2013}, design optimization \cite{do1,do2}, and optimal control \cite{oc}.

The computational challenges in high-fidelity simulations have led to the creation of reduced-order models (ROMs). ROMs aim to streamline the complex calculations of the full-order model (the detailed high-fidelity simulation) by reducing the complexity of the problem. While the accuracy of ROMs may slightly drop compared to the full-order-model, they can be considerably faster which makes them very attractive in situations where a small compromise in accuracy is acceptable. Many ROM methods are based on projecting data snapshots from the full-order model.

Methods like proper orthogonal decomposition (POD), the reduced basis method, and the balanced truncation method, which are linear projection techniques, have become increasingly popular in the field of reduced-order modeling (ROM) \cite{doi:10.1146/annurev.fl.25.010193.002543,rbm,Safonov1988ASM}. These approaches have demonstrated effectiveness in a range of applications, including fluid dynamics \cite{https://doi.org/10.48550/arxiv.2203.16494,doi:10.1137/19M1242963,Stabile_2018,https://doi.org/10.1002/num.21835,Copeland_2022,https://doi.org/10.48550/arxiv.2201.07335,MCLAUGHLIN20162407,math9141690}. Recently, non-linear projection techniques \cite{https://doi.org/10.48550/arxiv.2009.11990,https://doi.org/10.48550/arxiv.2011.07727,LEE2020108973,diaz2023fast}, particularly those using auto-encoders \cite{doi:10.1126/science.1127647,NIPS1992_cdc0d6e6}, have started to gain traction as an alternative, as these methods achieved higher performance in problems characterized by dominant advection \cite{https://doi.org/10.48550/arxiv.2009.11990,lasdi,glasdi}.

Projection-based ROM methods driven by data are broadly divided into two categories: intrusive and non-intrusive. Intrusive ROMs, which are informed by the underlying physics, need direct access to the governing equations \cite{rbm,diaz2023fast,https://doi.org/10.48550/arxiv.2011.07727,LEE2020108973,doi:10.1137/19M1242963,https://doi.org/10.1002/num.21835,Copeland_2022,math9141690,mcbane2022stress}. This aspect contributes to the sturdiness of their predictions and often requires a smaller amount of data from the full-order model (FOM). However, they also necessitate access to the FOM solver and specific details of implementation, like the discretized residual of the PDE. On the other hand, non-intrusive ROMs do not depend on the governing equations and are solely based on data-driven methods. These typically employ interpolation techniques to associate parameters with their respective ROM predictions \cite{osti_1420279,Marjavaara2006CFDDO,unknown,kutz_2017}. Being entirely black-box in nature, however, these methods often lack interpretability and robustness, sometimes facing challenges in accurate generalization and showing limitations in performance.

To address these issues, newer strategies are integrating projection methods with latent space dynamics learning. In these approaches, the latent space is considered as a dynamical system governed by a set of ordinary differential equations (ODEs). By precisely identifying these ODEs, it becomes possible to forecast the dynamics within the latent space and then map them back into the space of full-order solutions.

Several techniques have been developed to learn governing equations from data, including the popular method known as \textit{Sparse Identification of Non-Linear Dynamics} (SINDy) \cite{doi:10.1073/pnas.1517384113}. SINDy works by forming a collection of potential terms for the governing ordinary differential equations (ODEs) and uses linear regression to calculate the relevant coefficients. This approach has been widely used and laid the groundwork for the development of various SINDy-based algorithms including both regression-based \cite{doi:10.1126/sciadv.1602614,wsindy1,wsindy2,wendy,ident,weakident,doi:10.1098/rsos.211823} and neural network approaches \cite{https://doi.org/10.48550/arxiv.2211.10575,Chen_2021,BONNEVILLE2022100115,STEPHANY2022360}.

Champion et al. \cite{doi:10.1073/pnas.1906995116} introduced a method to identify sets of governing ODEs directly in the latent space of an auto-encoder. Although this method shows promise, the identified ODEs are not parameterized based on simulation parameters, which limits its applicability across different scenarios. Conversely, Bai and Peng suggested a similar technique \cite{https://doi.org/10.48550/arxiv.2106.09658}, but with a linear projection using proper orthogonal decomposition (POD), and they incorporated parameterization of the latent space ODEs. This enhancement allows for ROM predictions at any point in the parameter space. However, their approach faces challenges when applied to advection-dominated problems due to the limitations of POD.

More recently, a framework originally proposed by Fries et al. \cite{lasdi}, \textit{Latent Space Dynamics Identification} (LaSDI) was proposed. Building on the work of Champion et al. and Bai and Peng, the main idea of the LaSDI algorithm is to identify the sets of ODEs governing the auto-encoder latent space, each corresponding to one training data point from the full-order model (FOM). Then, the coefficients of each set of ODEs are interpolated with respect to the FOM parameters, which allows for estimating the set of latent space governing equations for any new parameter in the parameter space. Integrating the ODEs and passing them to the decoder allows to reconstruct the ROM prediction. This initial work laid the foundation for many follow-up LaSDI-based algorithms \cite{lasdi,glasdi,BONNEVILLE2024116535,bonneville2023datadriven,he2022certified,wlasdi}. Directly building on LaSDI \cite{lasdi}, \textit{Greedy}-LaSDI (gLaSDI) \cite{glasdi} introduced additional constrains to the auto-encoder training loss, enhancing the model's robustness. gLaSDI also introduced a novel active learning strategy, based on the PDE residual error (intrusive ROM), to acquire additional FOM data during training where it is the most needed. Subsequently, \textit{Gaussian Process} LaSDI (GPLaSDI) was also introduced \cite{BONNEVILLE2024116535}. GPLaSDI builds on the elements introduced in gLaSDI, but also employs Gaussian process for interpolating the latent space governing ODEs. This allows for a purely data-driven active learning strategy (non-intrusive ROM), as well as predicting meaningful confidence bounds over the ROM predictions. More recently, WLaSDI \cite{wlasdi}, short for \textit{Weak form}-LaSDI introduced a weak formulation to accurately learn the ODE dynamics of the latent space in the presence of substantial noise. Intriguingly, even when there is no noise, using WLaSDI yields a more accurate ROM in many cases. \textit{Thermodynamics}-LaSDI (tLaSDI) \cite{park2024tlasdi} was also proposed, and introduces physical constrains to enforce the first and second laws of thermodynamics within the latent space, for enhanced accuracy.

In this chapter, we provide a comprehensive review of LaSDI algorithms. First, We go through the key LaSDI building blocks (sections \ref{sec2.1}, \ref{sec:ae}, \ref{sec:sindy}, \ref{sec:loss}, \ref{sec:interp} and \ref{prediction}) and also introduce elements more specific to WLaSDI (section \ref{sec:wlasdi}), tLaSDI (section \ref{sec:tlasdi}), gLaSDI (section \ref{residual-sampling}) and GPLaSDI (section \ref{variance-sampling}). Finally, We present results demonstrating the performance of the different LaSDI methods on various examples (sections \ref{sec:1dburger}, \ref{sec:2nlh} and \ref{sec:3vlasov}). Note that in this chapter, the term LaSDI usually refers the set of different LaSDI algorithms, but may (when specified) also refer to the original LaSDI paper \cite{lasdi}.

\section{Latent Space Dynamics Identification}
\subsection{Governing Equation of Physical Systems}
\label{sec2.1}
In this chapter, we consider physical phenomenons described by governing PDEs with the following form:
\begin{equation}
\label{pde}
\begin{cases}
    \,\displaystyle\frac{\partial \mathbf{u}}{\partial t}=\mathbf{f}(\mathbf{u},t,x\,|\,\pmb{\mu})\hspace{0.75in}(t,x)\in[0,t_{\text{max}}]\times\Omega\\[5pt]
    \,\mathbf{u}(t=0,x\,|\,\pmb{\mu})=\mathbf{u}_0(x\,|\,\pmb{\mu})\hspace{0.35in}\pmb{\mu}\in\mathcal{D}\\
\end{cases}
\end{equation}
In Equation \eqref{pde}, the solution $\mathbf{u}$ may represent either a scalar or vector field over the time-space domain $[0,t_{\text{max}}]\times\Omega$. The spatial domain $\Omega$ can be of any dimension, and the differential operator $\mathbf{f}$ might include a mix of linear and/or non-linear combinations of spatial derivatives and source terms. The governing equations, along with their initial and boundary conditions, are influenced by a set of parameters, represented by a vector $\pmb{\mu}$. The parameter space is designated as $\mathcal{D}\subseteq\mathbb{R}^{n_\mu}$ and can have any dimension (for instance, a 2D-parameter space with $n_\mu=2$ is considered in subsequent sections). For a specific parameter vector $\pmb{\mu}^{(i)}\in\mathcal{D}$, it's assumed that we can access the corresponding discretized solution of Equation \eqref{pde}, denoted as $\mathbf{U}^{(i)}$. Here, $\mathbf{U}^{(i)}$ is a matrix composed of sequential snapshots $\mathbf{u}_n^{(i)}$ at each time step $n$, arranged as $\mathbf{U}^{(i)}=[\mathbf{u}_0^{(i)},\dots,\mathbf{u}_{N_t}^{(i)}]^\top\in\mathbb{R}^{(N_t+1)\times N_u}$. This solution is acquired either through a full-order model solver or an experiment. In this chapter, all the FOM solutions are compiled into a $3^\text{rd}$ order tensor dataset $\mathbf{U}\in\mathbb{R}^{N_\mu \times (N_t+1)\times N_u}$, where $N_t$ and $N_u$ are the number of time steps and degrees of freedom, respectively, and $N_\mu$ is the number of available FOM solutions. 

\subsection{Auto-encoders}
\label{sec:ae}
An auto-encoder \cite{doi:10.1126/science.1127647,Goodfellow-et-al-2016} is a type of neural network specifically tailored for compressing large datasets by diminishing their dimensions through nonlinear transformation. It comprises two interconnected neural networks: the encoder, labeled as $\phi_e$ and parameterized by $\pmb{\theta}_\text{enc}$, and the decoder, labeled as $\phi_d$ and parameterized by $\pmb{\theta}_\text{dec}$. The encoder processes an input data snapshot $\mathbf{u}_n^{(i)}\in\mathbb{R}^{N_u}$ and generates a compressed version $\mathbf{z}_n^{(i)}\in\mathbb{R}^{N_z}$ in a latent space. Here, $N_z$ denotes the latent dimension, i.e., the number of variables in the latent space. This number is chosen based on design preferences, typically ensuring $N_z<\!\!<N_u$. Similar to the matrix $\mathbf{U}^{(i)}$, we concatenate the latent representations at each time step into a matrix $\mathbf{Z}^{(i)}=[\mathbf{z}_{0}^{(i)},\dots,\mathbf{z}_{N_t}^{(i)}]^\top\in\mathbb{R}^{(N_t+1)\times N_z}$. The latent variables for each time step and for each parameter $\pmb{\mu}^{(i)}$ are compiled into a third-order tensor $\mathbf{Z}\in\mathbb{R}^{N_\mu\times (N_t+1)\times N_z}$, similar in structure to the tensor $\mathbf{U}$. The decoder takes each $\mathbf{z}_n^{(i)}$ as input and generates a reconstructed version of $\mathbf{u}_n^{(i)}$, denoted as $\hat{\mathbf{u}}_n^{(i)}$.
\begin{equation}
\label{ae}
\begin{aligned}
    &\mathbf{z}_n^{(i)}=\phi_e(\mathbf{u}_n^{(i)}\,|\,\pmb{\theta}_\text{enc})\\
    &\hat{\mathbf{u}}_n^{(i)}=\phi_d(\mathbf{z}_n^{(i)}\,|\,\pmb{\theta}_\text{dec})
\end{aligned}
\end{equation}
$\pmb{\theta}_\text{enc}$ and $\pmb{\theta}_\text{dec}$ are learned using a numerical optimization algorithm that minimizes the $L_2$ norm of the difference between the set of input solutions $\mathbf{U}$ and the set of reconstructed solutions $\hat{\mathbf{U}}$. The reconstruction loss is defined as:
\begin{equation}
\begin{aligned}
    \mathcal{L}_\text{AE}(\pmb{\theta}_\text{enc},\pmb{\theta}_\text{dec})&=|\!|\mathbf{U}-\hat{\mathbf{U}}|\!|_2^2\\
    &=\frac{1}{N_\mu}\sum_{i=1}^{N_\mu}\bigg(\frac{1}{N_t+1}\sum_{n=0}^{N_t}|\!|\mathbf{u}_n^{(i)}-\phi_d(\phi_e(\mathbf{u}_n^{(i)}\,|\,\pmb{\theta}_\text{enc})\,|\,\pmb{\theta}_\text{dec})|\!|_2^2\bigg)
\end{aligned}
\end{equation}
It should be noted that the auto-encoder acts as a non-linear data projection. Alternatively, a linear projection method, such as POD, may be employed. This option alleviates the cost of training an auto-encoder, but it may result in lower accuracy, especially for advection-dominated problems. A more detailed comparison between auto-encoders and POD is outlined in LaSDI \cite{lasdi} and WLaSDI \cite{wlasdi}.

\subsection{Identification of Latent Space Dynamics}
\label{sec:sindy}

The encoder's role is to compress high-dimensional physical data, such as solutions of PDEs that span both space and time, into a more compact array of discrete and abstract latent variables, defined over the time dimension. As a result, the latent space can effectively be viewed as a dynamical system, but now governed by ordinary differential equations (ODEs). This feature is a key element in LaSDI algorithms. For each discrete time step, the behavior of the latent variables within this latent space can be represented by a specific equation form. At each time step, the dynamics of the latent variables in the latent space can be described by an equation of the following form:
\begin{equation}
\label{ode}
    \dot{\mathbf{Z}}^{(i)}=\psi_{DI}(\mathbf{Z}^{(i)}\,|\,\pmb{\mu}^{(i)})
\end{equation}
where $\psi_{DI}$ represents a Dynamics Identification (DI) function governing the latent space dynamics, which can be defined as a system of ODEs and identified using the \textit{Sparse Identification of Nonlinear Dynamics} (SINDy) method \cite{doi:10.1073/pnas.1517384113}. SINDy involves creating a library, $\pmb{\Theta}(\mathbf{Z}^{(i)})\in\mathbb{R}^{(N_t+1)\times N_l}$, which is composed of $N_l$ linear and nonlinear potential terms that could be part of the ODEs. This method is based on the premise that the time derivatives of $\mathbf{Z}$, denoted as $\dot{\mathbf{Z}}$, can be represented as a linear combination of these selected terms. Consequently, the equation that describes the system's behavior on the right-hand side can be estimated in the following manner:
\begin{equation}
\label{ode2}
    \dot{\mathbf{Z}}^{(i)}\approx
    \pmb{\Theta}(\mathbf{Z}^{(i)})\cdot\mathbf{\Xi}^{(i)\top}
\end{equation}
where $\mathbf{\Xi}^{(i)}\in\mathbb{R}^{N_z\times N_l}$ denotes an ODE coefficient matrix associated with $\pmb{\mu}^{(i)}$. The selection of terms in $\pmb{\Theta}(\cdot)$ is a design choice. To capture the latent space dynamics more accurately, it may be desirable to include a broad variety of terms, although this may lead to sets of ODE more challenging and longer to solve numerically. $\pmb{\Theta}(\cdot)$ can be written explicitly as:
\begin{equation}
    \pmb{\Theta}(\mathbf{Z}^{(i)})=
    \begin{bmatrix}
    \mathbf{b}_0(\mathbf{z}_0^{(i)}) & \mathbf{b}_1(\mathbf{z}_0^{(i)}) & \mathbf{b}_2(\mathbf{z}_0^{(i)}) & \dots & \\
    \vdots & \vdots & \vdots\\
    \mathbf{b}_0(\mathbf{z}_{N_t}^{(i)}) & \mathbf{b}_1(\mathbf{z}_{N_t}^{(i)}) & \mathbf{b}_2(\mathbf{z}_{N_t}^{(i)}) & \dots &\\
    \end{bmatrix}_{(N_t+1) \times N_l}
\end{equation}
where $\mathbf{b}_q$ are basis functions that output arbitrary linear and/or non-linear transformations of each terms in $\mathbf{z}_n^{(i)}=[z_{n,1}^{(i)},\dots,z_{n,N_z}^{(i)}]$ to build the SINDy library. $z^{(i)}_{n,j}$ represents the $j^\text{th}$ latent variable for parameter $\pmb{\mu}^{(i)}$ at time step $n$. Sparse linear regressions are performed between $\pmb{\Theta}(\mathbf{Z}^{(i)})$ and $dz^{(i)}_{:,j}/dt$ for each $j\in[\![1,N_z]\!]$. The resulting coefficients associated with each SINDy term are stored in a vector $\pmb{\xi}_j^{(i)}=[\xi_{j,1}^{(i)},\dots,\xi_{j,N_l}^{(i)}]\in\mathbb{R}^{N_l}$. The system of SINDy regressions can be expressed as:
\begin{equation}
    \frac{dz^{(i)}_{:,j}}{dt}=\begin{bmatrix}
    \dot{z}_{0,j}^{(i)}\\
    \vdots\\
    \dot{z}_{N_t,j}^{(i)}
    \end{bmatrix}
    =\pmb{\Theta}(\mathbf{Z}^{(i)})\cdot\pmb{\xi}_j^{(i)\top}
    \label{ode3}
\end{equation}

Each set of coefficients $\pmb{\xi}_j^{(i)}$ is compiled into a coefficient matrix $\mathbf{\Xi}^{(i)}=[\pmb{\xi}_1^{(i)\top},\dots,\pmb{\xi}_{N_z}^{(i)\top}]^\top\in\mathbb{R}^{N_z\times N_l}$. Since a unique set of ODEs controls the dynamics within the latent space for each parameter vector $\pmb{\mu}^{(i)}\in\mathcal{D}$, several SINDy regressions are conducted simultaneously to identify the respective sets of ODE coefficients $\mathbf{\Xi}^{(i)}$, where $i\in[\![1,N_\mu]\!]$. The ensemble of ODE coefficient matrices $\mathbf{\Xi}=[\mathbf{\Xi}^{(1)},\dots,\mathbf{\Xi}^{(N_\mu)}]\in\mathbb{R}^{N_\mu \times N_z\times N_l}$, corresponding to each ODE system, is learned by minimizing the mean-squared-error loss specific to SINDy.
\begin{equation}
\label{sindy_loss}
\begin{aligned}
    \mathcal{L}_\text{SINDy}(\mathbf{\Xi})&=|\!|\dot{\mathbf{Z}}-\dot{\hat{\mathbf{Z}}}|\!|_2^2\\
    &=\frac{1}{N_\mu}\sum_{i=1}^{N_\mu}\bigg(\frac{1}{N_z}\sum_{j=1}^{N_z}\bigg|\!\bigg|\frac{dz_{:,j}^{(i)}}{dt}-\pmb{\Theta}(\mathbf{Z}^{(i)})\cdot\pmb{\xi}_j^{(i)\top}\bigg|\!\bigg|_2^2\bigg)
\end{aligned}
\end{equation}
The time derivatives $\dot{z}_{n,j}^{(i)}$ may be computed using a finite difference \cite{BONNEVILLE2024116535} or, alternatively, using the chain rule \cite{lasdi,glasdi}:
\begin{equation}
\label{der_z}
    \frac{\partial \mathbf{z}_n^{(i)}}{\partial t} = \frac{\partial \mathbf{z}_n^{(i)}}{\partial \mathbf{u}_n^{(i)}}
    \cdot\frac{\partial \mathbf{u}_n^{(i)}}{\partial t}=\nabla_\mathbf{u}\phi_e(\mathbf{u}_n^{(i)}\,|\,\pmb{\theta}_\text{enc})\cdot\mathbf{\dot{u}}_n^{(i)}
\end{equation} 
The aforementioned approach is a local DI method where each parameter has its own coefficient matrix. During testing, the local SINDy of training parameters can be exploited to estimate the SINDy associated with the testing parameter. More details will be presented in Section \ref{sec:interp}. Alternatively, a global DI approach can also be adopted in which only one coefficient matrix is employed to model the latent space dynamics for all parameters.
\subsection{Weak-form Identification}
\label{sec:wlasdi}
The coefficient matrix $\mathbf{\Xi}^{(i)}=[\pmb{\xi}_1^{(i)\top},\dots,\pmb{\xi}_{N_z}^{(i)\top}]^\top$ referenced in equation \eqref{ode2} can alternatively be determined using weak-form equation learning methods. This approach replaces the need to directly approximate pointwise derivatives from data (in the presence of noise, this is a particularly challenging task). Instead, the variance-reduction nature of the weak-form facilitates a more robust and accurate system recovery. In this section, we present WLaSDI \cite{wlasdi}, a direct extension of LaSDI \cite{lasdi} that leverages the weak-form equation learning technique. \\ \\
We begin by transforming equation \eqref{ode3} into the weak-form through multiplication by an absolutely continuous test function $\phi(t): \mathbb{R} \to \mathbb{R}$ and integrating it over the time domain: \\
\begin{equation*}
    \int_{t_a}^{t_b} \frac{dz^{(i)}_{:,j}}{dt} \cdot \phi(t)dt  =    \int_{t_a}^{t_b} \phi(t) \cdot \pmb{\Theta}(\mathbf{Z}^{(i)})\cdot\pmb{\xi}_j^{(i)\top}dt
\end{equation*}
Integration by parts yields: 
\begin{equation*}
    \phi(t_b)z^{(i)}_{t_b,j} - \phi(t_a)z^{(i)}_{t_a,j} -  \int_{t_a}^{t_b} z^{(i)}_{:,j} \cdot \phi'(t)dt = \int_{t_a}^{t_b} \phi(t) \cdot \pmb{\Theta}(\mathbf{Z}^{(i)})\cdot\pmb{\xi}_j^{(i)\top}dt
\end{equation*}
Choosing $\phi$ to be compactly supported in $(t_a, t_b)$, i.e., $ \phi(t_b) =  \phi(t_a) = 0$, we have:
\begin{equation} 
   -  \int_{t_a}^{t_b} z^{(i)}_{:,j} \cdot \phi'(t)dt = \int_{t_a}^{t_b} \phi(t) \cdot \pmb{\Theta}(\mathbf{Z}^{(i)})\cdot\pmb{\xi}_j^{(i)\top}dt.
    \label{ode_weakform}
\end{equation}
We use the trapezoidal rule to discretize the integrals in equation \eqref{ode_weakform} numerically and assume that observations of the system's state have a uniform time step of $\Delta t$.\textsuperscript{1}\footnote{\textsuperscript{1}It is possible to use a non-uniform time interval, but this necessitates modifications to the test function matrices.} Let $\{\phi_{k}\}_{m=1}^{N_k}$ be the set of compactly supported test functions placed uniformly along the time domain, we define the test function matrices: 
\begin{equation}\label{eq:tf_matrix}
\begin{array}{rl}
\mathbf{\Phi}_{kn} =  \Delta t \phi_k(t_n), \quad \mathbf{\Phi} \in \mathbb{R}^{N_k\times (N_t+1)}\\
\mathbf{\dot{\Phi}}_{kn} =  \Delta t \phi_k'(t_n), \quad \mathbf{\dot{\Phi}} \in \mathbb{R}^{N_k\times (N_t+1)} 
\end{array}
\end{equation}
From there, we have 
\begin{equation}\label{eq:G_b}
\begin{array}{rl}
\mathbf{G}^{(i)} & := \mathbf{\Phi}\pmb{\Theta}(\mathbf{Z}^{(i)}) \in\mathbb{R}^{N_k \times N_l }\\
\mathbf{b}_j^{(i)} & :=-(\dot{\mathbf{\Phi}}z_{:, j}^{(i)})^\top\in\mathbb{R}^{N_k}
\end{array}
\end{equation}
for $z_{:, j}^{(i)} = [z_{0,j}^{(i)},\cdots, z_{N_t,j}^{(i)}]^\top$ and $j\in[\![1,N_z]\!]$. By solving the least square problem, we identify the coefficient vector $\pmb{\xi}_j^{(i)\top}$ subject to: 
\begin{equation}
\underset{\pmb{\xi}_j^{(i)}\in\mathbb{R}^{N_l}}{\text{minimize}} \hspace{3pt} \left\Vert \mathbf{G}^{(i)}\pmb{\xi}_j^{(i)\top}-\mathbf{b}_j^{(i)}\right\Vert _{2}^{2}
\label{eq:WENDy}
\end{equation}
The weak-SINDy loss can then be defined as:
\begin{equation}
\label{weak_sindy_loss}
    \mathcal{L}_\text{SINDy}(\mathbf{\Xi})=\frac{1}{N_\mu}\sum_{i=1}^{N_\mu}\bigg(\frac{1}{N_z}\sum_{j=1}^{N_z}\left\Vert \mathbf{G}^{(i)}\pmb{\xi}_j^{(i)\top}-\mathbf{b}_j^{(i)}\right\Vert _{2}^{2}\bigg)
\end{equation}
The test functions employed in weak-form equation learning include high order polynomial or $C^\infty$ bump functions. For further details about how to choose these test functions, please refer to \cite{wendy,wsindy2,wsindy1}.

\subsection{Thermodynamics-informed Latent Space Dynamics Identification (tLaSDI)}
\label{sec:tlasdi}

tLaSDI \cite{park2024tlasdi} utilizes a neural network-based model to embed the \textit{General Equation for Non-Equilibrium Reversible-Irreversible Coupling} (GENERIC) formalism \cite{grmela1997dynamics,ottinger2005beyond} into the latent space dynamics.  
The GENERIC formalism is a mathematical framework that covers both conservative and dissipative systems, offering a general description of beyond-equilibrium thermodynamic systems. 
This formalism models dynamical systems through four key functions -- the scalar functions $E$ and $S$ that representing the system's total energy and entropy, respectively, and two matrix-valued functions, $L$ and $M$, referred to as the Poisson and friction matrices, respectively.

\begin{equation}
\label{eq:GENERIC}
\dot{\mathbf{z}} = L(\mathbf{z}) \nabla E(\mathbf{z}) +M(\mathbf{z}) \nabla S(\mathbf{z})\textrm{ s.t. }
\begin{cases}
L(\mathbf{z}) \nabla S(\mathbf{z}) = M(\mathbf{z}) \nabla E(\mathbf{z}) = \mathbf{0}\\[5pt]
L(\mathbf{z}) = -L(\mathbf{z})^\top\\[5pt]
M(\mathbf{z})\textrm{ is symmetric and positive semi-definite}
\end{cases}
\end{equation}

The degeneracy and structural conditions ensure the first and second laws of thermodynamics, i.e., the properties of energy conservation and non-decreasing entropy, 
$\frac{d}{dt}E(\mathbf{z}) = 0$ and $\frac{d}{dt}S(\mathbf{z}) \geq 0$.

\subsubsection{GFINNs}
Several neural network-based models were proposed to incorporate the GENERIC formalism \cite{hernandez2021structure,lee2021machine,zhang2022gfinns}.
While other alternatives are equally applicable, tLaSDI utilizes the model proposed in \cite{zhang2022gfinns}, namely, \textit{GENERIC Formalism Informed Neural Networks} (GFINNs).
We direct readers to refer to \cite{zhang2022gfinns} for more details.
The GFINNs consists of $4$ neural networks -- $E_{\text{NN}}$, $S_{\text{NN}}$, $L_{\text{NN}}$ and $M_{\text{NN}}$ -- that represent $E$, $S$, $L$ and $M$ in equation \eqref{eq:GENERIC}, respectively. 
These neural networks are constructed to exactly satisfy the degeneracy conditions in \eqref{eq:GENERIC} while also being sufficiently expressive to capture the underlying dynamics from data.
We denote the GFINNs by $\phi_{\text{G}}(\, \cdot \, |\,\pmb{\theta}_{\text{G}})$, where $\pmb{\theta}_{\text{G}}$ represents the neural network parameters in the model. 

We note that GFINNs are global DI model and do not have parameter dependence. 
Instead, tLaSDI lets the network parameters of the encoder and the decoder depend on parameters by employing hypernetworks \cite{ha2017hypernetworks}.

The hypernetwork is a neural network that takes a parameter $\pmb{\mu}^{(i)}$ as input and outputs the network parameters of another neural network. 
In tLaSDI, two distinct hypernetworks $\phi_{e}^{\text{hyp}}(\,\cdot \, |\pmb{\theta}_{\text{enc}}^{\text{hyp}})$ and $\phi_{d}^{\text{hyp}}(\, \cdot \, |\pmb{\theta}_{\text{dec}}^{\text{hyp}})$ are utilized to output the network parameters of the encoder and decoder.
The network parameters for encoder and decoder at parameter $\pmb{\mu}^{(i)}$ are denoted as 
$\pmb{\theta}_\text{enc}^{(i)} := \phi_{e}^{\text{hyp}}(\pmb{\mu}^{(i)}|\pmb{\theta}_{\text{enc}}^{\text{hyp}})$ and
$\pmb{\theta}_\text{dec}^{(i)} := \phi_{d}^{\text{hyp}}(\pmb{\mu}^{(i)}|\pmb{\theta}_{\text{dec}}^{\text{hyp}})$, respectively.

In tLaSDI, the GFINNs can be trained by minimizing the \textit{model} and \textit{integration} losses, respectively $\mathcal{L}_\text{MOD}$ and $\mathcal{L}_\text{INT}$:
\begin{equation}
\begin{aligned}
    \mathcal{L}_\text{MOD}(\pmb{\theta}_\text{enc}^{\text{hyp}},\pmb{\theta}_\text{dec}^{\text{hyp}},\pmb{\theta}_\text{G}) = &\frac{1}{N_\mu (N_t+1)}\sum_{i=1}^{N_\mu}\sum_{n=0}^{N_t} \bigg(\left\Vert \dot{\mathbf{z}}^{(i)}_n  - \phi_{\text{G}}(\mathbf{z}^{(i)}_n\,|\,\pmb{\theta}_{\text{G}})
     \right\Vert^2_2\\
     &+\left\Vert \dot{\mathbf{u}}^{(i)}_n  - \nabla_\mathbf{z} \phi_d (\mathbf{z}_{n}^{(i)}\,|\,\pmb{\theta}_{\text{dec}}^{(i)}) \phi_{\text{G}}(\mathbf{z}^{(i)}_n\,|\,\pmb{\theta}_{\text{G}})
     \right\Vert^2_2 \bigg)
     \end{aligned}
\end{equation}
\begin{equation}
\label{integ_loss}
\mathcal{L}_\text{INT}(\pmb{\theta}_\text{enc}^{\text{hyp}},\pmb{\theta}_\text{G})=\frac{1}{N_\mu N_t}\sum_{i=1}^{N_\mu }\sum_{n=0}^{N_t-1} \left\Vert \mathbf{z}_{n+1}^{(i)}-\mathbf{z}_n^{(i)} - 
     \int_{t_n}^{t_{n+1}} \phi_{\text{G}}(\mathbf{z}^{(i)}\,|\,\pmb{\theta}_{\text{G}}) dt\right\Vert^2_2
\end{equation}
The derivative $\dot{\mathbf{z}}^{(i)}_n $ is computed using equation (\ref{der_z}), and the integral in the integration loss (equation \eqref{integ_loss}) is approximated by employing a numerical integrator (e.g., Runge-Kutta methods).

\subsection{Training Loss}
\label{sec:loss}
In LaSDI \cite{lasdi}, the auto-encoder and SINDy are trained separately. Hower, for enhanced accuracy and robustness when identifying the latent space dynamics, in gLaSDI \cite{glasdi}, the auto-encoder and SINDy are trained simultaneously. gLaSDI also introduced an additional loss term: the mean-squared-error of the velocity, computed through the chain rule:
\begin{equation}
\label{chainrule2}
    \dot{\hat{\mathbf{u}}}^{(i)}_n=\frac{\partial \hat{\mathbf{u}}_n^{(i)}}{\partial \mathbf{z}_n^{(i)}}
    \cdot\frac{\partial \mathbf{z}_n^{(i)}}{\partial t}=\nabla_\mathbf{z}\phi_d(\phi_e(\mathbf{u}_n^{(i)}\,|\,\pmb{\theta}_\text{enc})\,|\,\pmb{\theta}_\text{dec})\cdot\pmb{\Theta}(\mathbf{Z}^{(i)})\cdot\mathbf{\Xi}^{(i)\top}
\end{equation}
\begin{equation}
    \mathcal{L}_\text{VEL}(\pmb{\theta}_\text{enc},\pmb{\theta}_\text{dec},\mathbf{\Xi})=|\!|\dot{\mathbf{U}}-\dot{\hat{\mathbf{U}}}|\!|_2^2
\end{equation}
In GPLaSDI \cite{BONNEVILLE2024116535}, a penalty is also added to the SINDy coefficients. This enforce smaller values of the ODE coefficients, which leads to better conditioned ODEs. The LaSDI training loss, weighted with hyperparameters $\beta_1$, $\beta_2$, $\beta_3$ and $\beta_4$ is:
\begin{equation}
\label{loss}
\begin{aligned}
    \mathcal{L}(\pmb{\theta}_\text{enc},\pmb{\theta}_\text{dec},\mathbf{\Xi})&=\beta_1\mathcal{L}_\text{AE}(\pmb{\theta}_\text{enc},\pmb{\theta}_\text{dec})+\beta_2\mathcal{L}_\text{SINDy}(\mathbf{\Xi})\\[5pt]
    &+\beta_3\mathcal{L}_\text{VEL}(\pmb{\theta}_\text{enc},\pmb{\theta}_\text{dec},\mathbf{\Xi})+\beta_4|\!|\mathbf{\Xi}|\!|_2^2
\end{aligned}
\end{equation}
where $\mathcal{L}_\text{SINDy}$ may be either the \textit{vanilla} SINDy loss (equation \eqref{sindy_loss}) or the weak SINDy loss (equation \eqref{weak_sindy_loss}). In tLaSDI, the latent space dynamics is not learned through SINDy, so the training loss has a slightly different (yet equivalent) formalism:
\begin{equation}
\label{loss_tLaSDI}
\begin{aligned}
    \mathcal{L}(\pmb{\theta}_\text{enc}^{\text{hyp}},\pmb{\theta}_\text{dec}^{\text{hyp}},\pmb{\theta}_\text{G})&=\lambda_1\mathcal{L}_\text{AE}(\pmb{\theta}_\text{enc}^{\text{hyp}},\pmb{\theta}_\text{dec}^{\text{hyp}})
    +\lambda_2\mathcal{L}_\text{MOD}(\pmb{\theta}_\text{enc}^{\text{hyp}},\pmb{\theta}_\text{dec}^{\text{hyp}},\pmb{\theta}_\text{G})\\[5pt]
    &+\lambda_3\mathcal{L}_\text{INT}(\pmb{\theta}_\text{enc}^{\text{hyp}},\pmb{\theta}_\text{G})
    +\lambda_4\mathcal{L}_\text{JAC}(\pmb{\theta}_\text{enc}^{\text{hyp}},\pmb{\theta}_\text{dec}^{\text{hyp}}) 
\end{aligned}
\end{equation}
where $\lambda_1$, $\lambda_2$, $\lambda_3$ and $\lambda_4$ are hyperparameters (analog to $\beta_1$, $\beta_2$, $\beta_3$ and $\beta_4$ in equation \eqref{loss}).
The first and second terms of $\mathcal{L}_\text{MOD}$ are analog to $\mathcal{L}_\text{SINDy}$ and $\mathcal{L}_\text{VEL}$ in the gLaSDI training loss, respectively.
Furthermore, the \textit{Jacobian} loss $\mathcal{L}_\text{JAC}$ is defined as
\begin{equation}
\begin{aligned}
      \mathcal{L}_\text{JAC}(\pmb{\theta}_\text{enc}^{\text{hyp}},\pmb{\theta}_\text{dec}^{\text{hyp}}) 
    & = \frac{1}{N_\mu (N_t+1)}\sum_{i=1}^{N_\mu}\sum_{n=0}^{N_t} \left\Vert \bigg( I  - \nabla_\mathbf{z} \phi_d (\mathbf{z}_{n}^{(i)}|\pmb{\theta}_{\text{dec}}^{(i)})
    \nabla_\mathbf{u} \phi_e (\mathbf{u}^{(i)}_n|\pmb{\theta}_{\text{enc}}^{(i)}) \bigg) \dot{\mathbf{u}}^{(i)}_n 
     \right\Vert^2_2
     \end{aligned}
\end{equation}
The loss components are based on the error estimates of ROM approximation provided in \cite{park2024tlasdi}.


\subsection{Parameterized Latent Space Dynamics Interpolation}
\label{sec:interp}
The auto-encoder learns a latent space dynamical representation of each training parameter. In the local Dynamics Identification (DI) approach, a set of SINDy's or weak-SINDy's learns the ODEs governing the latent space dynamics of training parameters. However, the set of ODE coefficients associated with a new parameter $\pmb{\mu}^{(*)}$ (distinct from the set of training parameters $\pmb{\mu}^{(i)}\in\mathcal{D}$) remains unknown. It can be estimated by finding an interpolant $\psi:\pmb{\mu}^{(*)}\mapsto\mathbf{\Xi}^{(*)}$ of the learned ODE coefficients, given $\pmb{\mu}^{(i)}\in\mathcal{D}$ and $\mathbf{\Xi}^{(i)}$, $i\in[\![1,N_\mu]\!]$. Several interpolants have been proposed in the literature, and we briefly cover three of them here, RBF interpolation, as used in LaSDI \cite{lasdi}, $k-$NN, as used in gLaSDI \cite{glasdi}, and Gaussian process regression (GP), as introduced in GPLaSDI \cite{BONNEVILLE2024116535}.

\subsubsection{RBF Interpolation}

Let us consider a domain $\mathcal{D}_\text{DI}\subseteq\mathcal{D}$. $\mathcal{D}_\text{DI}$ may be a subset of the parameter space (local Dynamics Identification), or the full space (global Dynamics Identification). The interpolated ODE coefficient matrix $\mathbf{\Xi}^{(*)}$ can be expressed as:
\begin{equation}
\mathbf{\Xi}^{(*)}=\sum_{\pmb{\mu}^{(i)}\in\mathcal{D}_\text{DI}}w_i\psi_i(d(\pmb{\mu}^{(*)},\pmb{\mu}^{(i)}))
\end{equation}
$\psi_i$ is a radial basis function (RBF). For example in LaSDI \cite{lasdi}, a multiquadric function is used:
\begin{equation}
\psi_i(d(\pmb{\mu}^{(*)},\pmb{\mu}^{(i)}))=\sqrt{\frac{d(\pmb{\mu}^{(*)},\pmb{\mu}^{(i)})^2}{\epsilon}+1}
\end{equation}
$d(\pmb{\mu}^{(*)},\pmb{\mu}^{(i)})$ is the euclidian distance between $\pmb{\mu}^{(*)}$ and $\pmb{\mu}^{(i)}$, $\epsilon$ is a length-scale typically chosen as the average distance between each $\pmb{\mu}^{(i)}\in\mathcal{D}_\text{DI}$ and the coefficients $w_i$ are computed by solving a linear system depending on the set of matrices $\mathbf{\Xi}^{(i)}$ associated with each parameter $\pmb{\mu}^{(i)}\in\mathcal{D}_\text{DI}$.

\subsubsection{$k-$Nearest-Neighbours Interpolation} 

Similarly to RBF interpolation, in $k-$NN interpolation, the ODE coefficient matrix can be written as:
\begin{equation}
\mathbf{\Xi}^{(*)}=\sum_{i\in\mathcal{N}_k(\pmb{\mu}^{(*)})}\psi_i(\pmb{\mu}^{(*)},\pmb{\mu}^{(i)})\mathbf{\Xi}^{(i)}
\end{equation}
$\mathcal{N}_k(\pmb{\mu}^{(*)})$ is the set of indices of the $k$-nearest-neighbours of $\pmb{\mu}^{(*)}$ (i.e. associated with $k$ training parameters $\pmb{\mu}^{(i)}$). The interpolation basis $\psi_i$ are defined as:
\begin{equation}
    \psi_i(\pmb{\mu}^{(*)},\pmb{\mu}^{(i)})=\frac{|\!|\pmb{\mu}^{(*)}-\pmb{\mu}^{(i)}|\!|_M^{-2}}{\sum_{j\in\mathcal{N}_k(\pmb{\mu}^{(*)})}|\!|\pmb{\mu}^{(*)}-\pmb{\mu}^{(j)}|\!|_M^{-2}}
\end{equation}
where $|\!|\cdot|\!|_M$ is the Mahalanobis distance. Here, partition of unity is satisfied ($\sum_i\psi_i(\pmb{\mu}^{(*)},\pmb{\mu}^{(i)})=1$) and convexity preservation is guaranteed \cite{glasdi}. 

\subsubsection{Gaussian Process Regression}

In a Gaussian process regression \cite{books/lib/RasmussenW06}, the fitting function is stochastic and is assumed to follow a prior Gaussian probability distribution:
\begin{equation}
    \psi\sim\mathcal{N}(0,k(\pmb{\mu}, \pmb{\mu}))
\end{equation}
$k(\cdot, \cdot)$ is an arbitrary covariance kernel, function of the input data (i.e. the set of parameters $\pmb{\mu}^{(i)}$) and some trainable hyper-parameters. In GPLaSDI \cite{BONNEVILLE2024116535}, the kernel is chosen as a squared-exponential function. Using both Bayes' Rule and the sum and product rules of probability, it can be shown that the predictive distribution of the interpolated set of ODE coefficients, $\mathbf{\Xi}^{(*)}$, follows a Gaussian distribution:
\begin{equation}
\label{preddistrib}
p(\mathbf{\Xi}^{(*)}|\pmb{\mu}^{(*)},\pmb{\mu},\mathbf{\Xi})=\mathcal{N}(\mathbf{\Xi}^{(*)}|m^{(*)},s^{(*)2}) 
\end{equation}
$m^{(*)}$ and $s^{(*)}$ are the predictive mean and standard deviation, which depend on $\pmb{\mu}^{(*)}$, $\pmb{\mu}$, $\mathbf{\Xi}$, and the kernel function. It should be emphasized that unlike RBF and $k-$NN regression, $\mathbf{\Xi}^{(*)}$ is here not deterministic and may thus take an infinity of possible values. Since it follows a Gaussian distribution, we may only consider its most likely value, the mean ($m^{(*)}$), with some confidence interval defined by $s^{(*)}$. 

Note that in practice, it may be cumbersome to perform the GP inference over the entire matrix of coefficients $\mathbf{\Xi}^{(*)}$ altogether. Instead, we train a separate GP for each ODE coefficient term in $\mathbf{\Xi}^{(*)}$. More details on GP regression and inference can be found in \cite{BONNEVILLE2024116535,books/lib/RasmussenW06}.

\subsection{Predicting Solutions}
\label{prediction}
Once the auto-encoder has been properly trained and the latent space governing ODEs have been learned (section \ref{sec:ae}, \ref{sec:sindy}, \ref{sec:wlasdi} and \ref{sec:loss}), ROM prediction for an arbitrary test parameter $\pmb{\mu}^{(*)}$ is straightforward and can be made through the following steps:
\begin{enumerate}
    \item Estimate $\mathbf{\Xi}^{(*)}$ by interpolating the set of learned ODE coefficients using one of the interpolator introduced in section \ref{sec:interp}. If using GP regression, in-lieu of $\mathbf{\Xi}^{(*)}$, a sample from the predictive distribution may be used: $\mathbf{\Xi}^{(d)}\sim\mathcal{N}(\mathbf{\Xi}^{(*)}|m^{(*)},s^{(*)2})$, $d\in[\![1,N_s]\!]$, with $N_s$ an arbitrary number of samples.
    \item Build and solve the set of ODEs associated with $\pmb{\mu}^{(*)}$ (equation \eqref{ode2}) using a standard numerical integrator. The initial condition $\mathbf{z}_0^{(*)}$ can be computed using the encoder ($\mathbf{z}_0^{(*)}=\phi_e(\mathbf{u}_0^{(*)}\,|\,\pmb{\theta}_\text{enc})$), where $\mathbf{u}_0^{(*)}$ is a function of $\pmb{\mu}^{(*)}$. The approximate latent-space solution is noted $\tilde{\mathbf{Z}}^{(*)}$.
    \item Make a forward pass through the decoder to map the ROM prediction back into the full-order space. $\tilde{\mathbf{U}}^{(*)}=\phi_d(\tilde{\mathbf{Z}}^{(*)}\,|\,\pmb{\theta}_\text{dec})$
\end{enumerate}
Note that if using GP regression to interpolate the latent space governing equations, step 1, 2 and 3 may be repeated for a finite number of samples. The ROM prediction can then be taken as the average prediction over all the samples ($\tilde{\mathbf{U}}^{(*)}\equiv\mathbb{E}[\tilde{\mathbf{U}}^{(*)}]$). Similarly, the standard deviation $\mathbb{V}[\tilde{\mathbf{U}}^{(*)}]^{1/2}$ can also be computed to estimate prediction uncertainty. In tLaSDI, the prediction do not involve interpolation methods as GFINN is a global DI model independent of parameters. The tLaSDI prediction for a test parameter $\pmb{\mu}^{(*)}$ can be conducted through the following analog steps:
\begin{enumerate}
    \item Evaluate the network parameters of the encoder and decoder using corresponding hypernetworks, i.e., $\pmb{\theta}_\text{enc}^{(*)} := \phi_{e}^{\text{hyp}}(\pmb{\mu}^{(*)}|\pmb{\theta}_{\text{enc}}^{\text{hyp}})$
    and  $\pmb{\theta}_\text{dec}^{(*)} := \phi_{d}^{\text{hyp}}(\pmb{\mu}^{(*)}|\pmb{\theta}_{\text{dec}}^{\text{hyp}})$.
    \item Solve the latent space dynamics (equation \eqref{eq:GENERIC}) represented by GFINNs using a numerical integrator. The initial condition is given by  $\mathbf{z}_0^{(*)}=\phi_e(\mathbf{u}_0^{(*)}\,|\,\pmb{\theta}_\text{enc}^{(*)})$. 
    The approximate solution is denoted as $\tilde{\mathbf{Z}}^{(*)}$.
    \item Use the decoder to return the prediction in the full-order space, i.e., $\tilde{\mathbf{U}}^{(*)}=\phi_d(\tilde{\mathbf{Z}}^{(*)}\,|\,\pmb{\theta}_\text{dec}^{(*)})$.
\end{enumerate}

\section{Greedy Sampling}

One key aspect of LaSDI algorithms is to train an auto-encoder and learn the governing dynamics in the latent space. To that end, choosing the appropriate training data is crucial. In LaSDI \cite{lasdi} and WLaSDI \cite{wlasdi}, each FOM solution data point is associated with equispaced parameters $\pmb{\mu}^{(i)}$, located on a grid within the parameter space $\mathcal{D}$. While this approach is simple, easy to implement, and may provide satisfactory performance, more sophisticated approaches have been proposed in gLaSDI \cite{glasdi} and GPLaSDI \cite{BONNEVILLE2024116535}. In these two papers, an active learning strategy is employed, where instead of having access to the entire training dataset upfront, the training data is collected on the fly. This allows for collecting only the data that is the most needed while maximizing model performances. In this section, we present the key aspects of these two methods. The general greedy sampling framework is described in algorithm \ref{alg:cap}

\subsection{Residual-Based Greedy Sampling (gLaSDI)}
\label{residual-sampling}
For a given test parameter, the ROM prediction error may be estimated using the time-averaged PDE residual of the FOM governing equation:
\begin{equation}
\label{res}
    e_\text{RES}(\tilde{\mathbf{U}}^{(*)})=\frac{1}{N_{ts}}\sum_{n=1}^{N_{ts}}|\!|\mathbf{r}(\tilde{\mathbf{u}}^{(*)}_n,\tilde{\mathbf{u}}^{(*)}_{n-1}|\pmb{\mu}^{(*)})|\!|_2
\end{equation}
$\mathbf{r}$ refers to the discretized residual of the governing PDE (equation \eqref{pde}). $N_{ts}$ is the number of time steps used to estimate the residual. When $\mathbf{r}$ is expensive to evaluate, it may be desirable to evaluate the residual only at a handfull of time steps, and we may thus consider $N_{ts}\ll N_t$. 

In gLaSDI \cite{glasdi}, the training process is initialized using a limited number ($N_\mu$) of training FOM simulations. The training is then paused every $N_{up}$ epochs, and the (current) LaSDI model is used to make predictions at a finite number of test parameters $\pmb{\mu}^{(*)}\in\mathcal{D}^h$ ($\mathcal{D}^h$ is a discretized representation of the parameter space). Each prediction $\tilde{\mathbf{U}}^{(*)}$ is plugged into equation \eqref{res} to compute the residual error, and the next sampling parameter can be selected as the one associated with the ROM solution yielding the largest residual:
\begin{equation}
\label{rbgs}
    \pmb{\mu}^{(N_\mu+1)}=\arg\max_{\pmb{\mu}^{(*)}\in\mathcal{D}^h}\big[e_\text{RES}(\tilde{\mathbf{U}}^{(*)})\big]
\end{equation}
A high-fidelity simulation can then be run for parameter $\pmb{\mu}^{(N_\mu+1)}$, and the resulting solution is added to the training dataset before resuming the training process. This operation may be repeated until the budget for running FOM simulations has been exhausted, or the ROM predictions have become sufficiently accurate.
The residual-based greedy sampling is also employed in tLaSDI \cite{park2024tlasdi} following the algorithm proposed in \cite{glasdi}.
 
\subsection{Uncertainty-Based Greedy Sampling (GPLaSDI)}
\label{variance-sampling}
As described in section \ref{prediction}, when using GP interpolation, the uncertainty in the latent space dynamics may be propagated through the decoder, and the ROM prediction variance can be computed for any time step $n$:
\begin{equation}
\label{rom_var}
    \mathbb{V}[\tilde{\mathbf{u}}_n^{(*)}]=\frac{1}{N_s}\sum_{d=1}^{N_s}(\phi_\text{dec}(\tilde{\mathbf{z}}_n^{(*,d)}|\pmb{\theta}_\text{dec})-\mathbb{E}[\tilde{\mathbf{u}}_n^{(*)}])^2
\end{equation}
$\tilde{\mathbf{z}}_n^{(*,d)}$ represent the latent space dynamics solved for a sampled ODE coefficient matrix $\mathbf{\Xi}^{(d)}$ (equation \eqref{preddistrib}). The variance at each time step can be computed and the variance across time and space is written $\mathbb{V} [\tilde{\mathbf{U}}^{(*)}]=[\mathbb{V}[\tilde{\mathbf{u}}_0^{(*)}],\dots,\mathbb{V}[\tilde{\mathbf{u}}_{N_t}^{(*)}]]$.

An uncertainty based strategy for picking the next sampling parameter $\pmb{\mu}^{(N_\mu+1)}$ can now easily be implemented. The general approach is very similar to the residual-based greedy sampling strategy presented in section \ref{residual-sampling}, but instead of picking the parameter yielding the largest residual error, we pick the parameter yielding the largest ROM variance (at any point in time and space):
\begin{equation}
\label{vbgs}
\pmb{\mu}^{(N_\mu+1)}=\arg\max_{\pmb{\mu}^{(*)}\in\mathcal{D}^h}\big[\max_{(t,x)}\mathbb{V}[\tilde{\mathbf{U}}^{(*)}]\big]
\end{equation}
Note that in order to properly quantify the ROM variance ($\mathbb{V}[\tilde{\mathbf{U}}^{(*)}]$), for each $\pmb{\mu}^{(*)}\in\mathcal{D}^h$, the GP predictive distribution needs to be sampled $N_s$ times, and the corresponding ODE needs to be solved and fed to the decoder. Therefore, this approach requires more ROM prediction than the residual-based greedy sampling approach. However, this approach does not require to evaluate the PDE residual at any time. Thus, GPLaSDI is non-intrusive and may be particularly suitable when the residual is difficult to implement, expensive to evaluate, or simply unknown (e.g. legacy codes for high-fidelity simulations).

\begin{algorithm}[!h]
\caption{Auto-encoder Training with Greedy Sampling}\label{alg:cap}
\begin{algorithmic}[1]
\While{$h<N_{epoch}$}
    \State Compute $\mathbf{Z}=\phi_\text{enc}(\mathbf{U}\,|\,\pmb{\theta}_\text{enc})$ and $\hat{\mathbf{U}}=\phi_\text{dec}(\mathbf{Z}\,|\,\pmb{\theta}_\text{dec})$ (eq. \eqref{ae})
    \State Compute $\mathcal{L}(\pmb{\theta}_\text{enc},\pmb{\theta}_\text{dec}, \mathbf{\Xi})$,  and update $\pmb{\theta}_\text{enc}$, $\pmb{\theta}_\text{dec}$, and $\mathbf{\Xi}$ using gradient descent 
    \If{$h\mod N_{up}\equiv 0$}
        \State Interpolate $\mathbf{\Xi}$ using RBF, $k-$NN or GP interpolation (section \ref{sec:interp})
        \For{$\pmb{\mu}^{(*)}\in\mathcal{D}^h$}
            \If{Residual-Based Greedy Sampling}
                \State Compute $\tilde{\mathbf{U}}^{(*)}$ and $e_\text{RES}(\tilde{\mathbf{U}}^{(*)})$ (eq. \eqref{res})
            \EndIf
            \If{Variance-Based Greedy Sampling}
            \State Compute $\mathbb{V}[\tilde{\mathbf{U}}^{(*)}])$ (eq. \eqref{rom_var})
            \EndIf
        \EndFor
        \State Find $\displaystyle\pmb{\mu}^{(N_\mu+1)}$ using eq. \eqref{rbgs} or eq. \eqref{vbgs}
        \State Collect $\mathbf{U}^{(N_\mu+1)}$ using FOM solver
        \State Update dataset $\mathbf{U}=[\mathbf{U}^{(1)},\dots,\mathbf{U}^{(N_\mu)},\mathbf{U}^{(N_\mu+1)}]$ and $N_\mu=N_\mu+1$
    \EndIf
    \State Update $h=h+1$
\EndWhile
    
\end{algorithmic}
\end{algorithm}

\section{Application}
We now briefly summarize results and examples originally introduced in \cite{lasdi,glasdi,wlasdi,BONNEVILLE2024116535,park2024tlasdi}, and cover in particular the 1D Burgers equation (section \ref{sec:1dburger}), the non-linear heat equation (section \ref{sec:2nlh}), and a two-stream plasma instability problem (section \ref{sec:3vlasov}). In each of the following section, the error metric used is the maximum relative error defined as:
\begin{equation}
    e(\tilde{\mathbf{U}}^{(*)},\mathbf{U}^{(*)})=\max_n\bigg(\frac{|\!|\tilde{\mathbf{u}}^{(*)}_n-\mathbf{u}^{(*)}_n|\!|_2}{|\!|\mathbf{u}^{(*)}_n|\!|_2}\bigg),
\end{equation}
where the quantities with a tilde are predictions.

\subsection{1D Burgers Equation}
\label{sec:1dburger}
In this first application, we use a simple benchmark problem to compare the different LaSDI algorithms introduced in the previous two sections. We consider the following inviscid 1D Burgers equation:
\begin{equation}
    \begin{cases}
    \displaystyle\frac{\partial u}{\partial t}+u\frac{\partial u}{\partial x}=0\hspace{0.3in}(t,x)\in[0,1]\times[-3,3]\\[10pt]
    \displaystyle u(t,x=3)=u(t,x=-3)
    \end{cases}
\end{equation}
The initial condition is parameterized by $\pmb{\mu}=\{a,w\}\in\mathcal{D}$, and the parameter space is defined as $\mathcal{D}=[0.7,0.9]\times[0.9,1.1]$:
\begin{equation}
    u(t=0,x\,|\,\pmb{\mu})=a\exp\bigg(-\frac{x^2}{2w^2}\bigg)\hspace{0.3in}\pmb{\mu}=\{a,w\}\\
\end{equation}
The parameter space is discretized into a square grid $\mathcal{D}^h$ with a stepping of $\Delta a=\Delta w=0.01$, resulting in a total of 441 grid points ($21$ values in each dimension). The high-fidelity solver employs a classic backward Euler time integration, and a finite difference discretization in space ($\Delta x=6\cdot10^{-3}$ and $\Delta t=10^{-3}$). 

\subsubsection{LaSDI vs. gLaSDI} We first compare the performances of LaSDI trained with pre-selected data, or data selected on the fly using residual-based greedy sampling (gLaSDI) \cite{glasdi}. For the encoder, we employ a 1001-100-5 fully-connected hidden layers/hidden units architecture ($N_z=5$), with sigmoid activation functions, and a symmetric architecture for the decoder. The loss hyperparameters are taken as $\beta_1=1$, $\beta_2=0.1$, $\beta_3=0.1$ and $\beta_4=0$. We also consider $50,000$ training epochs, with a $N_{up}=2000$ greedy sampling rate. The SINDy library is restricted to linear terms and the ODE coefficients are interpolated with $k-$NN interpolation ($k=1$ and $k=3$ are employed during gLaSDI training and evaluation, respectively, and $k=4$ is employed for LaSDI). The four corner parameters of $\mathcal{D}$ are used to generate the initial training data points ($\pmb{\mu}^{(1)}=\{0.7,0.9\}$, $\pmb{\mu}^{(2)}=\{0.9,0.9\}$, $\pmb{\mu}^{(3)}=\{0.7,1.1\}$ and $\pmb{\mu}^{(4)}=\{0.9,1.1\}$) At the end of the training, a total of 25 training data points are being used.

\begin{figure}[!h]
\hspace{-1.cm}
    \includegraphics[width=1.1\textwidth]{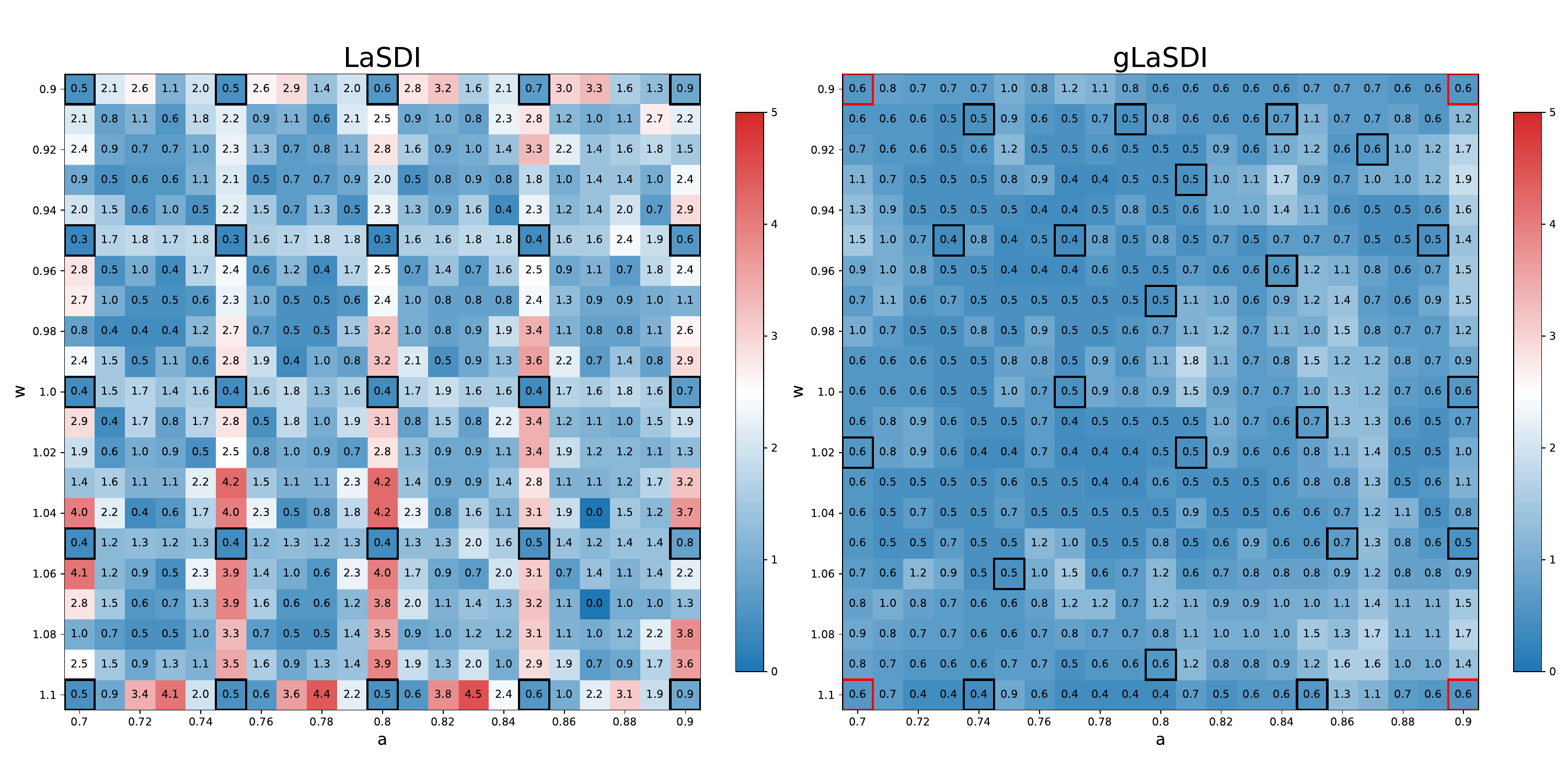}
    \caption{1D Burgers Equation -- Maximum relative error ($\%$) using LaSDI on a pre-selected uniform training grid, and using gLaSDI (residual-based) active learning. The black boxes represent the parameters corresponding to the training FOM datapoints (and the red boxes represent the four initialization training points, in the gLaSDI case).}
    \label{1db_fig1}
\end{figure}

Figure \ref{1db_fig1} shows a comparison between LaSDI and gLaSDI maximum relative error across the parameter space. Both achieve excellent accuracy: at most $4.5\%$ error for LaSDI (trained on a $5\times5$ parameter grid), and at most $1.9\%$ error for gLaSDI. Since gLaSDI selects training data based on the worst residual error, areas of the parameter space yielding large prediction errors will eventually concentrate more data, helping to capture the correct physics more accurately and faster. 

\subsubsection{GPLaSDI vs. gLaSDI} We now compare the performance of variance-based greedy sampling (GPLaSDI) and residual-based greedy sampling (gLaSDI). We consider a similar auto-encoder architecture as in the previous experiment (1001-100-5-100-1001), and use $\beta_1=1$, $\beta_2=0.1$, $\beta_3=0$ and $\beta_4=10^{-6}$ for the loss hyperparameters. Like earlier, we restrict the SINDy library to linear terms only. Since GPLaSDI employs Gaussian processes for interpolating the latent space ODEs, we also interpolate the latent space ODEs in gLaSDI with GPs for a more fair comparison. Even if GPs are employed along with gLaSDI here, the uncertainty information is not exploited and the additional training data is selected solely through the residual error (as in the previous experiment). In GPLaSDI, we employ $N_s=20$ ODE samples to compute the ROM prediction variance, and train both models for $N_{epoch}=28,000$ epochs with sampling rate $N_{up}=2000$ (13 training data points are collected on the fly, along with the 4 initialization corner points).

\begin{figure}[!h]
\hspace{-1.cm}
    \includegraphics[width=1.1\textwidth]{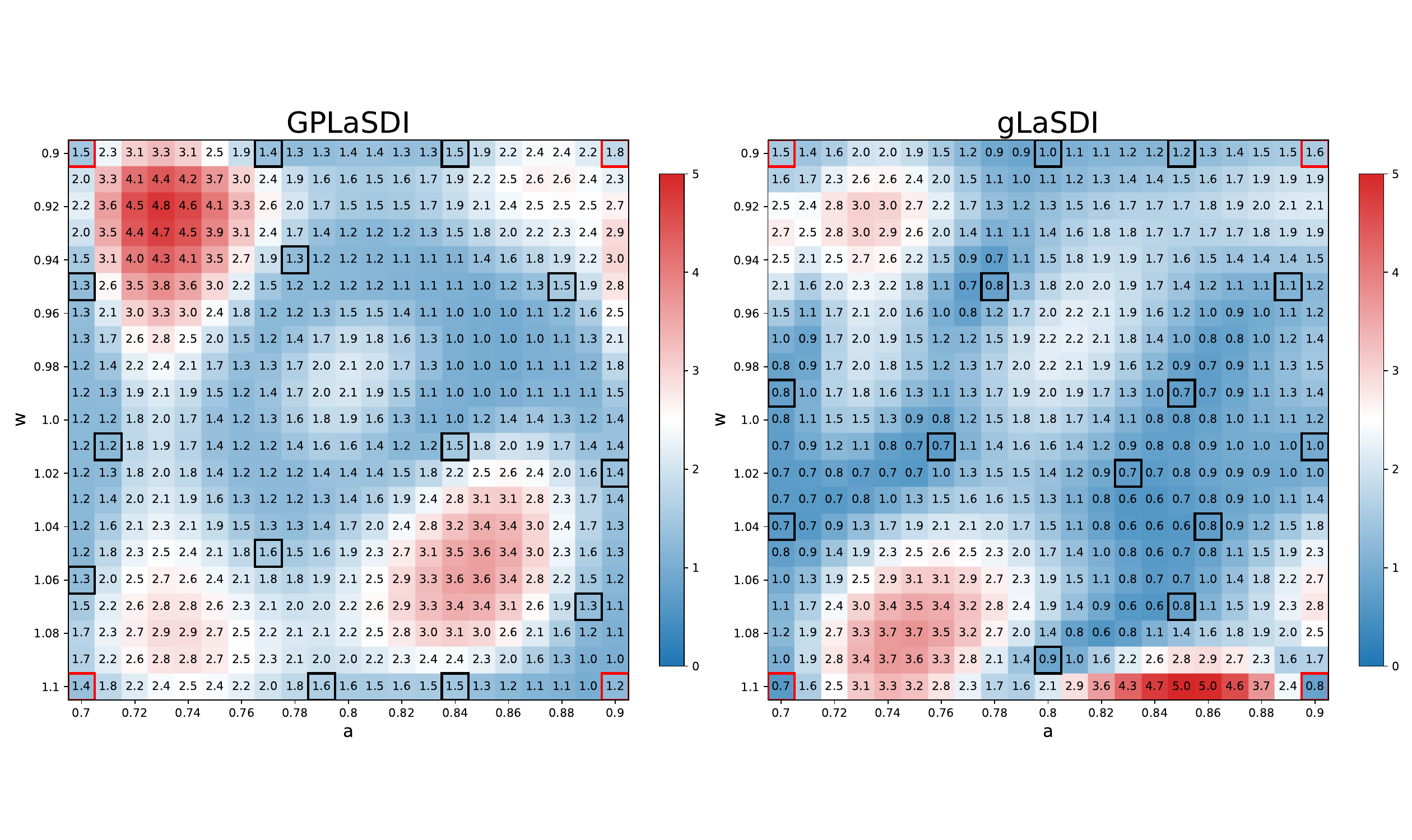}
    \caption{1D Burgers Equation -- Maximum relative error ($\%$) using GPLaSDI and using gLaSDI}
    \label{1db_fig2}
\end{figure}

\begin{figure}[!h]
\centering
    \includegraphics[width=0.6\textwidth]{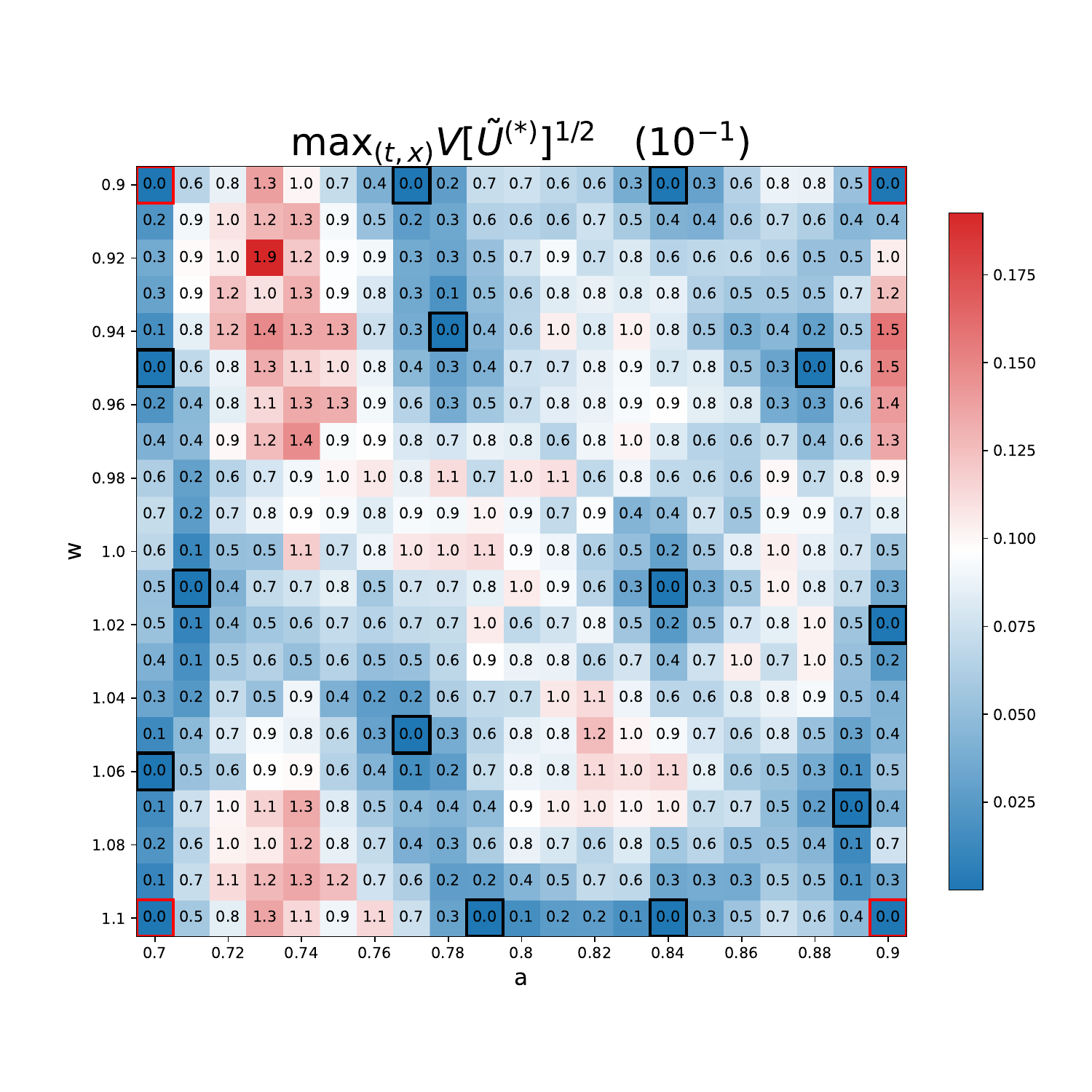}
    \caption{1D Burgers Equation -- Maximum Predictive Variance}
    \label{1db_var}
\end{figure}

Figure \ref{1db_fig2} shows a comparison between GPLaSDI and gLaSDI. Both cases achieve very similar performance (at most $4.8\%$ and $5\%$ error for GPLaSDI and gLaSDI respectively). One might expect gLaSDI to perform generally better than GPLaSDI, because the former has direct access to physics information (through the PDE residual), while GPLaSDI is purely data-driven. Here however, GPLaSDI is able to achieve similar performance than gLaSDI. This indicates that in the present case, sampling new data for parameters yielding the largest ROM variance or the largest residual error is equivalent. Figure \ref{1db_var} shows the maximum ROM standard deviation across time and space, for each parameter of the discretized parameter space. There is a strong correlation between GPLaSDI's maximum relative error and maximum standard deviation, which is consistent with the expectation that reducing the prediction uncertainty should also reduce the error. 

\subsubsection{LaSDI vs. Weak-LaSDI}
\begin{figure}
\hspace{-1.cm}
    \includegraphics[width=1.1\textwidth]{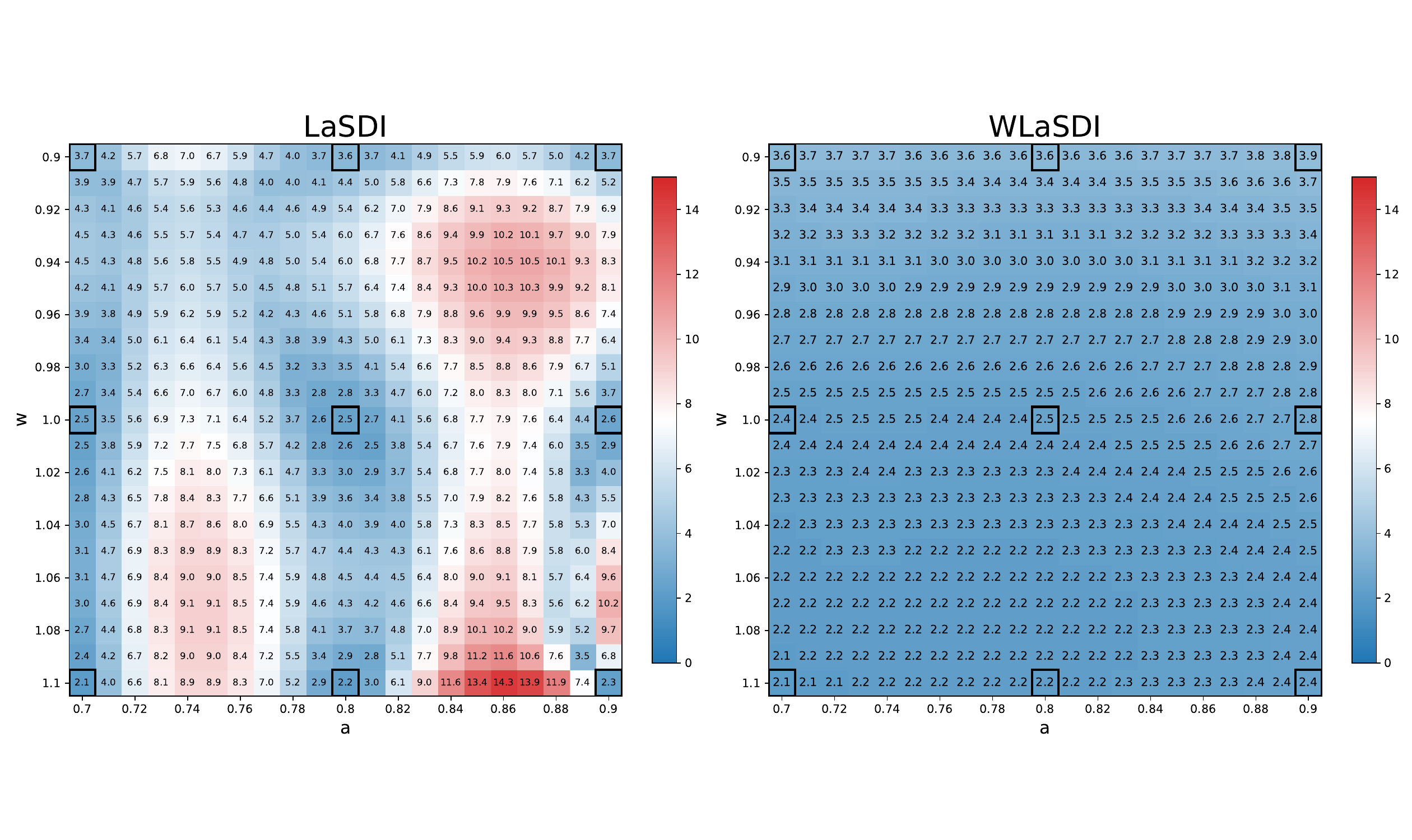}
    \caption{1D Burgers Equation -- Maximum relative error ($\%$) using LaSDI and using WLaSDI (no noise case). Note that this error is bounded below by the autoencoder projection error, which is about $3\%$ across the parameter space in this case.}
    \label{1db_fig4}
\end{figure}

\begin{figure}
\hspace{-1.cm}
    \includegraphics[width=1.1\textwidth]{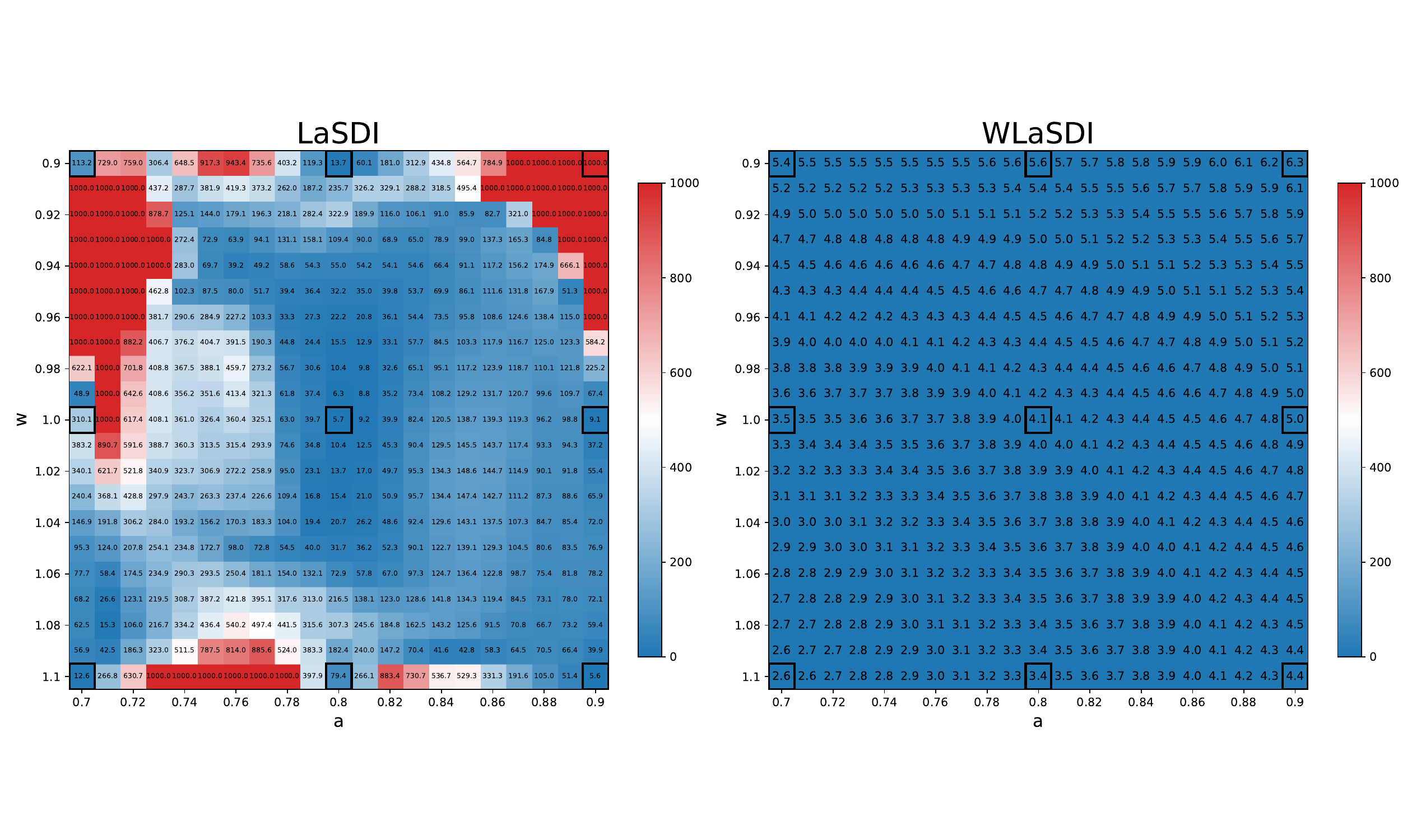}
    \caption{1D Burgers Equation -- Maximum relative error ($\%$) using LaSDI and using WLaSDI with $10\%$ added noise. The relative error is bounded below by the autoencoder projection error, which is about $5\%$ across the parameter space.}
    \label{1db_fig5}
\end{figure}

In this experiment, we evaluate the performance of LaSDI and WLaSDI, comparing them under a fixed number of training data points. Active learning is intentionally excluded, as the primary goal is to demonstrate the robustness of the weak-form in comparison with the strong form. A shallow-masked auto-encoder architecture is utilized \cite{KimChoiWidemannEtAl2022JournalofComputationalPhysics} for the nonlinear ROM creation. The training data contains 9 high-fidelity simulations (of the inviscid Burgers' equation) corresponding to the combination of each value of $a$ and $w$ where $a = [0.7, 0.8, 0.9]$ and $w = [0.9, 1.0, 1.1]$. We consider 2 cases: Figure \ref{1db_fig4} illustrates the difference between LaSDI and WLaSDI without any added noise, whereas Figure \ref{1db_fig5} demonstrates the difference between LaSDI and WLaSDI with $10\%$ added Gaussian white noise.  The ODE coefficients governing the latent space dynamics are interpolated using Radial basis function interpolation. To ensure a fair comparison, the type of radial basis function and the radius are kept consistent between LaSDI and WLaSDI. Note that in both cases, the relative error is bounded below by the error of the autoencoder. Exploring better autoencoder architecture may lead to improvements. In the noise-free case, the projection error introduced by the autoencoder is about $3\%$ across the parameter space. Meanwhile, for the noisy case, it is about $5\%$.

For Figure \ref{1db_fig4}, the depicted results show the maximum relative error across the parameter space when employing LaSDI and WLaSDI without any noise. In the case of LaSDI, the error varies between $3\%$ and $14\%$, while WLaSDI demonstrates a lower error, remaining below $4\%$ (nearly achieving the minimum possible error, given the autoencoder architecture limitations). Figure \ref{1db_fig5} depicts the same comparison, but for data with $10\%$ noise. In the noisy scenario, LaSDI frequently returns values with errors in excess of $1000\%$ (any value exceeding $1000\%$ is capped at $1000\%$ in Figure \ref{1db_fig5}), whereas WLaSDI maintains a more controlled error, staying below $7\%$. In these instances, the strong form requires approximately $0.03$ seconds, while the weak-form computation takes about $0.3$ seconds. Therefore, for a tenfold increase in computation time, the error is reduced by two to three orders of magnitude (at the $10\%$ noise level).

The perturbations in the data introduced by the noise propagate through the encoder, making it challenging to approximate the time derivative of the latent space. By eliminating the need to approximate the pointwise derivative as well as leveraging the variance-reduction nature of the integral, the weak form allows for more accurate identification of the governing ODEs. 

Given its robustness against noise, WLaSDI is particularly well suited for building numerical models based solely on experimental data\textsuperscript{2} \footnote{\textsuperscript{2} Note that the weak version of SDI methods would also be effective in attempting to find a ROM from models with a stochastic component, like a Particle-In-Cell simulation.} especially when combined with GPLaSDI for efficient data sampling. 


%

\subsubsection{tLaSDI vs. gLaSDI} In this last experiment, we compare the performance of tLaSDI and gLaSDI utilizing different temporal and parameter domains, specifically $(t,x)\in[0,2]\times[-3,3]$, $\mathcal{D}=[0.7,0.8]\times[0.9,1.0]$. Due to the extended temporal domain, the solutions exhibit stiff behaviors as time approaches the final time ($t=2$).
The training data are discretized with the spatial and time spacing $\Delta x=6/200$, $\Delta t=2/200$, which are subsampled from the high-fidelity data with $\Delta x=6\cdot10^{-3}$ and $\Delta t=2\cdot 10^{-3}$. We used the auto-encoder architecture (201-100-10-100-201) for both methods. 
The sigmoid and ReLU activation functions are used to train auto-encoders of gLaSDI and tLaSDI respectively. 
In tLaSDI, all neural networks within GFINNs consist of $5$ layers and $40$ neurons in each hidden layer, and the hyperbolic tangent activation function is employed to train them. 
The loss hyperparameters for gLaSDI are set to $\beta_1=1$, $\beta_2=1$, $\beta_3=1$ and $\beta_4=0$. For tLaSDI, the loss weights $\lambda_1=10^{-1}$, $\lambda_2=10^{-7}$, $\lambda_3=1$ and $\lambda_4=10^{-9}$ are employed. Both methods utilize the residual-based greedy sampling for $42,000$ training epochs with a sampling rate $N_{up}=2000$. A total of $21$ training data points are collected on the fly, in addition to $4$ initial samples at the corners of parameter space. For gLaSDI, quadratic polynomials are used for the SINDy library, and the ODE coefficients are interpolated with $k-$NN interpolation with $k=1$ and $k=5$ during training and evaluation, respectively. 

\begin{figure}[!h]
\hspace{-1.cm}
    \includegraphics[width=1.1\textwidth]{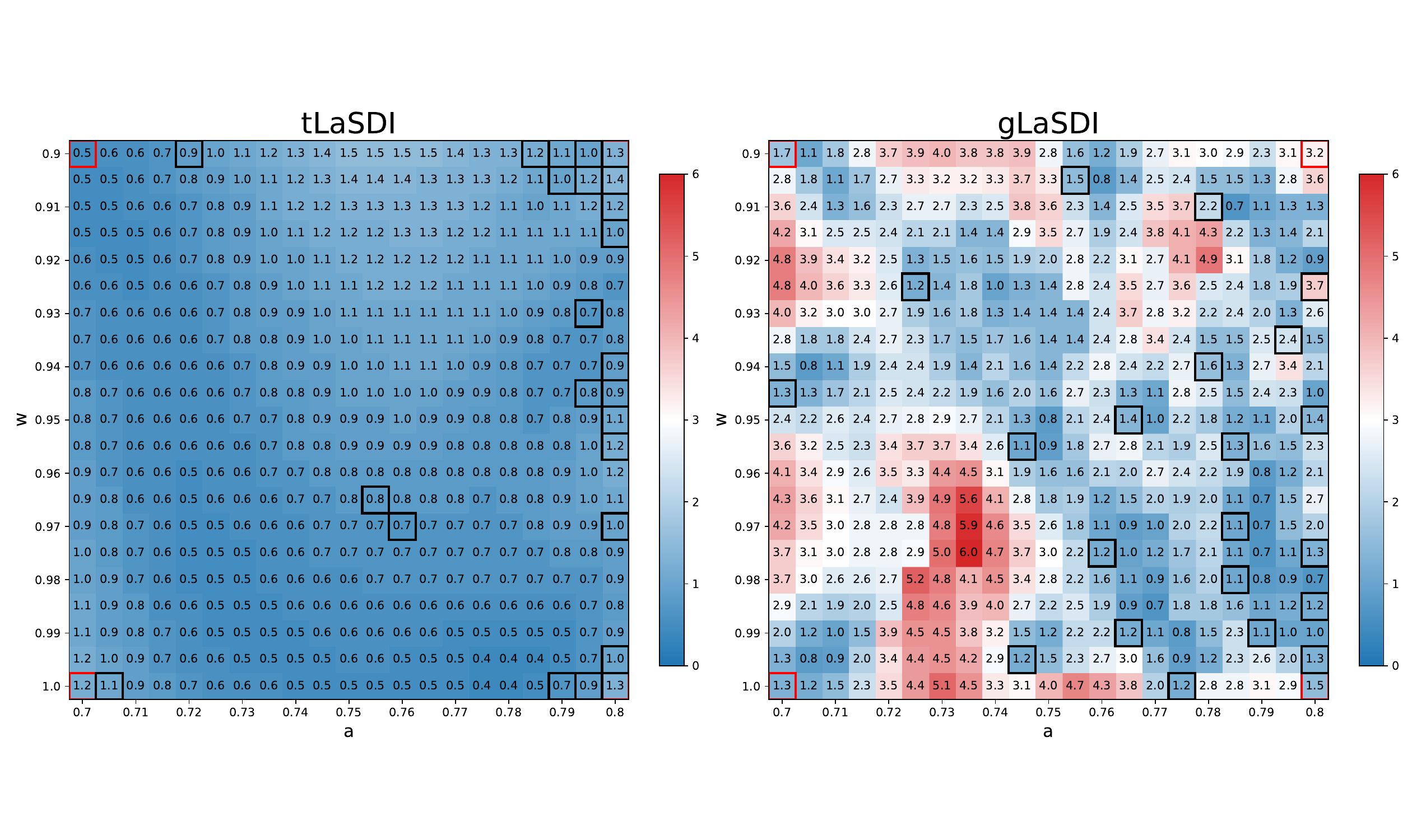}
    \caption{1D Burgers Equation -- Maximum relative error ($\%$) using tLaSDI and using gLaSDI}
    \label{1db_fig3}
\end{figure}

\begin{figure}[!h]
\hspace{-1.cm}
    \includegraphics[width=1.1\textwidth]{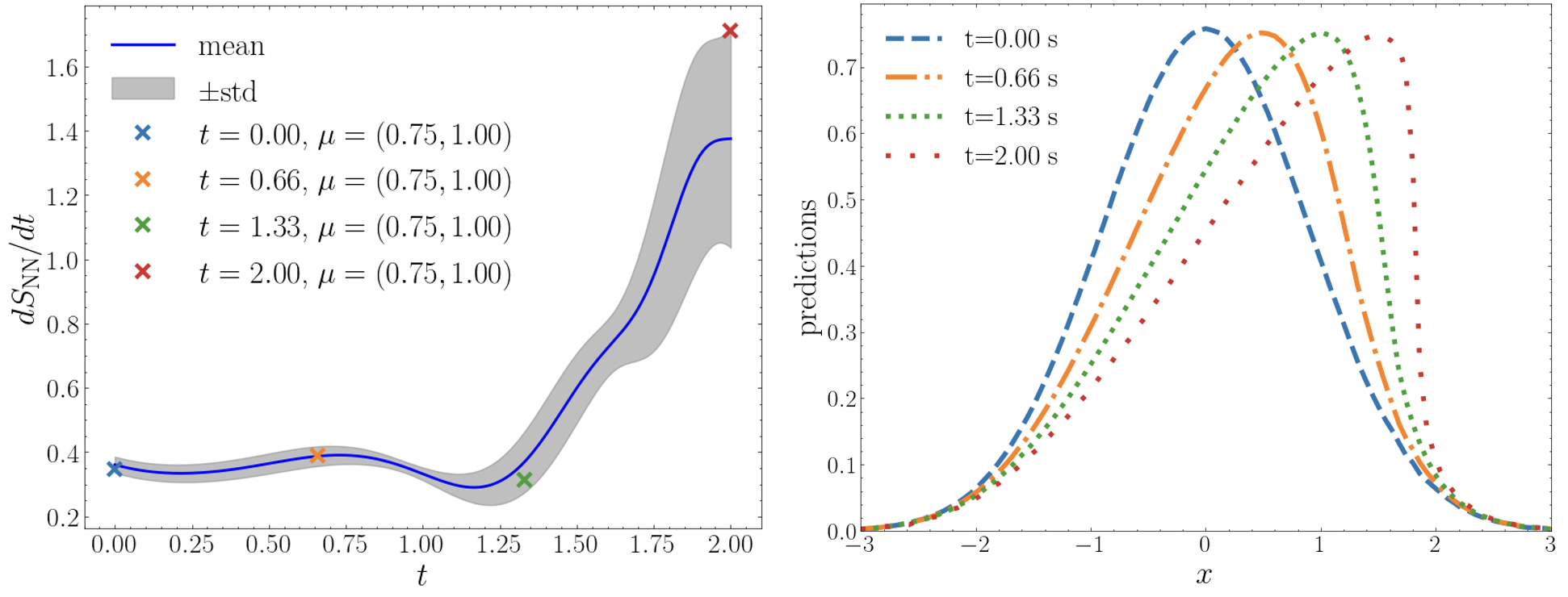}
    \caption{1D Burgers Equation --  Left: The mean and one standard deviation away from the mean of $\frac{d}{dt}S_\text{NN}$ across the test parameters. Right: The solution prediction by tLaSDI at varying times at $\mu = (0.75,1.00)$ whose corresponding $\frac{d}{dt}S_\text{NN}$ are marked in the left figure.}
    \label{1db_fig3-1}
\end{figure}

Figure \ref{1db_fig3} depicts a comparison between tLaSDI and gLaSDI in terms of maximum relative errors over the parameter domain. 
Both algorithms achieve high prediction accuracy (at most $1.5\%$ for tLaSDI and $6.0\%$ for gLaSDI). The sampled parameters for tLaSDI tend to cluster near the domain boundaries.
For both methods, a higher concentration of training data is observed in regions of large amplitudes ($a$) that correspond to the solutions exhibiting stiffer behavior.  
This demonstrates that both methods effectively select training data points using the greedy sampling strategy to accurately predict solutions exhibiting stiff patterns.
The effective prediction performance of tLaSDI is due to its loss function and incorporation of laws of physics (thermodynamics) in the latent space.    


tLaSDI constructs a thermodynamic structure in the latent space dynamics which provides the neural network based entropy function $S_{\text{NN}}$ in the latent space.
Figure~\ref{1db_fig3-1} illustrates the entropy production rate $\frac{d}{dt}S_{\text{NN}}(\mathbf{z}(t))$ with respect to time.
The mean over all the test parameters is reported and the shared area represents one standard deviation away from the mean. 
As promised by GFINNs, we observe that the entropy increases as the rate is always non-negative. 
The entropy production rate has the largest value at the final time $t=2$, which happens to be the time where the full-state solution exhibits the stiffest pattern.
See the right of Figure~\ref{1db_fig3-1} for the tLaSDI prediction at varying times.

\subsection{2D Non-Linear Heat Equation with Residual-based Greedy Sampling}
\label{sec:2nlh}

In this second example, we consider the non-linear heat equation in 2D introduced in gLaSDI \cite{glasdi}:

\begin{equation}
\begin{cases}
    \displaystyle\frac{\partial u}{\partial t}=\nabla\cdot(\kappa+\alpha u)\nabla u\hspace{0.3in}
    (t,x,y)\in[0,0.3]\times[0,1]\times[0,1]\\[15pt]
    \displaystyle\frac{\partial u}{\partial \mathbf{n}}=\mathbf{0}\hspace{0.3in}(x,y)\in\partial\Omega
\end{cases}
\end{equation}
The diffusion coefficients are chosen as $\kappa=0.5$ and $\alpha=0.01$. The initial condition is parameterized by $\pmb{\mu}=\{a,w\}\in\mathcal{D}$, and the parameter space is defined as $\mathcal{D}=[1,1.4]\times[4,4.3]$:
\begin{equation}
    u(t=0,x,y\,|\,\pmb{\mu})=a\sin\Big(w\sqrt{x^2+y^2}\Big)+a\hspace{0.3in}\pmb{\mu}=\{a,w\}
\end{equation}
The parameter space is discretized into a square grid $\mathcal{D}^h$ with stepping $\Delta a=0.02$ and $\Delta w=0.015$, resulting in a total of 441 grid points ($21$ values in each dimension). To generate the high-fidelity data, we rely on a finite element code, MFEM \cite{mfem}, and we discretize the spatial domain with 1024 first-order square elements. For time integration, we employ a backward Euler scheme ($\Delta t=5\cdot10^{-3}$). 

The encoder employs a 1089-100-3 fully-connected hidden layers/hidden units architecture ($N_z=3$), with sigmoid activation functions, and a symmetric architecture for the decoder. The loss hyperparameters are taken as $\beta_1=1$, $\beta_2=10^{-4}$, $\beta_3=10^{-4}$ and $\beta_4=0$. We also consider $2.5\cdot10^6$ training epochs, with a $N_{up}=10^5$ greedy sampling rate. The ODE coefficients are interpolated with $k-$NN interpolation ($k=1$ and $k=3$ are employed during gLaSDI training and evaluation, respectively, and $k=4$ is employed for LaSDI). The four corner parameters of $\mathcal{D}$ are used to generate the initial training data points ($\pmb{\mu}^{(1)}=\{1,4\}$, $\pmb{\mu}^{(2)}=\{1,4.3\}$, $\pmb{\mu}^{(3)}=\{1.4,4\}$ and $\pmb{\mu}^{(4)}=\{1.4,4.3\}$) At the end of the training, there are a total of 25 training data points. We compare the performance of gLaSDI (residual-based greedy sampling) with LaSDI trained on a $5\times5$ uniform parameter grid.

\begin{figure}[!h]
\hspace{-1.cm}
    \includegraphics[width=1.1\textwidth]{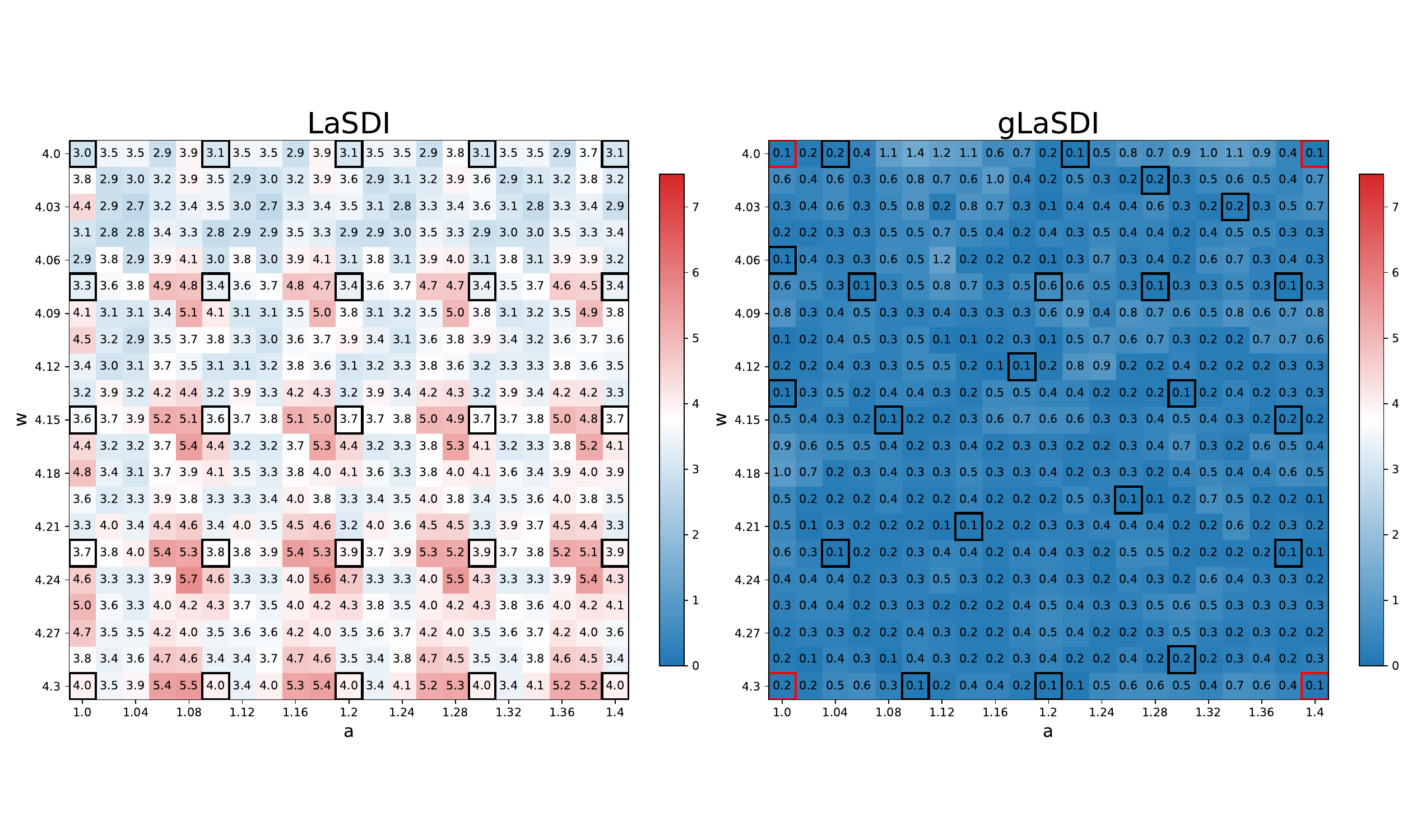}
    \caption{2D Non-Linear Heat Equation  -- Maximum relative error ($\%$) with a SINDy dictionnary restricted to linear terms, using LaSDI and using gLaSDI}
    \label{2dh_lin}
\end{figure}

\begin{figure}[!h]
\hspace{-1.cm}
    \includegraphics[width=1.1\textwidth]{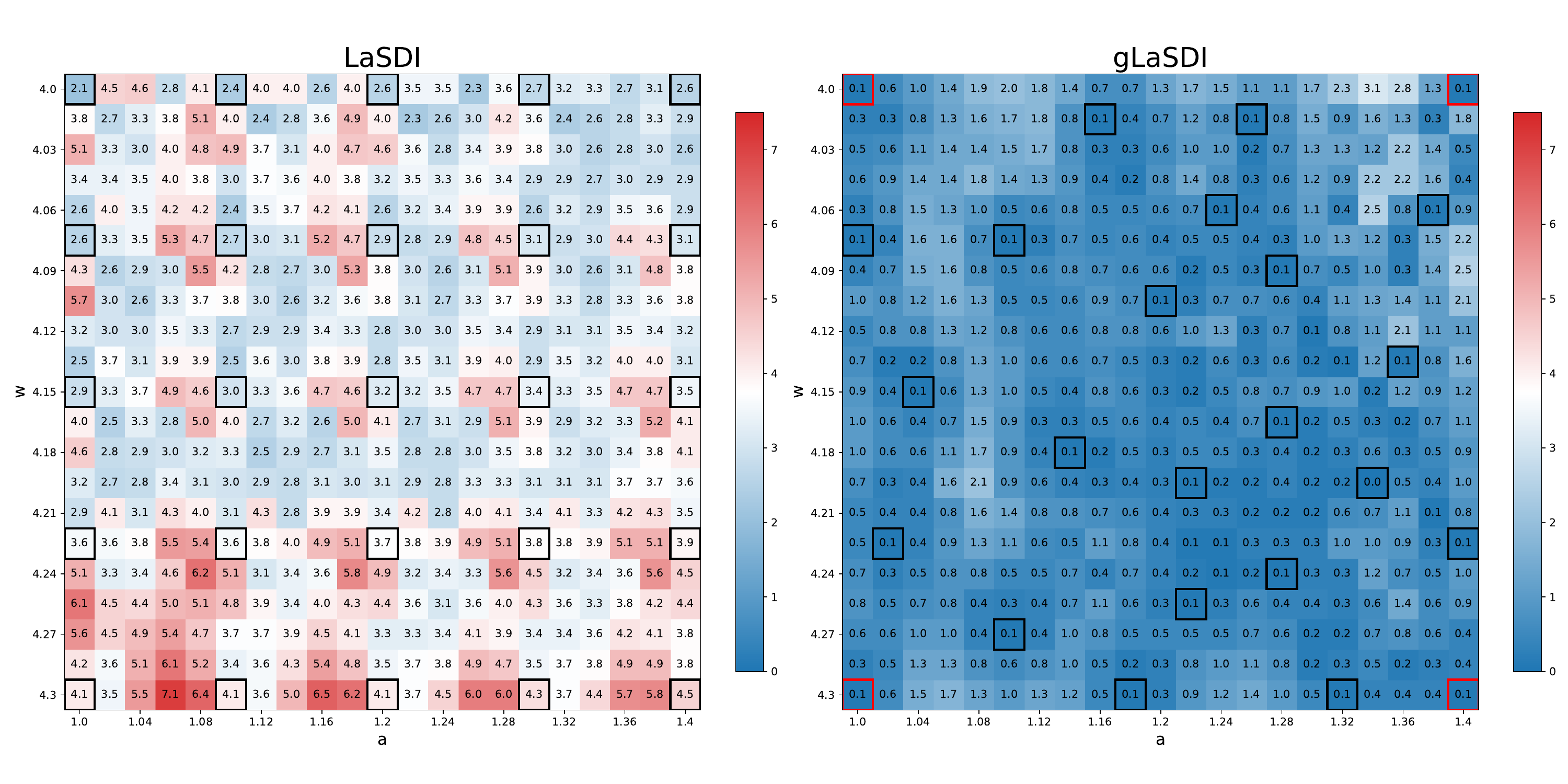}
    \caption{2D Non-Linear Heat Equation  -- Maximum relative error ($\%$) with a SINDy dictionnary restricted to linear and quadratic terms, using LaSDI and using gLaSDI}
    \label{2dh_quad}
\end{figure}

Figure \ref{2dh_lin} illustrates the maximum relative error when using LaSDI and gLaSDI, with a SINDy library restricted to linear terms only. As already observed with the 1D Burgers equation example, gLaSDI clearly outperforms LaSDI by selecting the most appropriate training data. gLaSDI achieves at most $1.2\%$ maximum relative error, whereas LaSDI maximum relative error can be as high as $5.7\%$. The choice of candidate terms used in the SINDy library is purely arbitrary. Thus, we do not have to restrict it to linear terms only. Figure \ref{2dh_quad} shows the maximum relative error for the same experiment, but this time using both linear and quadratic candidate terms in the SINDy library. Interestingly, a slight deterioration of performances is noticed (at most $3.1\%$ and $7.1\%$ error for gLaSDI and LaSDI, respectively). This indicates that while a broader selection of SINDy candidate terms may theoretically capture the latent space dynamics more accurately, it is not always the case in practice. A simpler set of candidate (e.g. linear terms) may sometime be sufficient. In fact, increasing the number of SINDy candidate terms means that more ODE coefficients need to be interpolated, so the potential for interpolation error increases accordingly. gLaSDI achieves $17\times$ speed-up (with linear and quadratic SINDy library), and $58\times$ speed-up (with linear SINDy library only). As a result, restricting the SINDy library to linear terms not only achieves better performance in this example, it is also faster. This is not surprising since in the later case, less ODE coefficients need to be interpolated, and the ODE are simpler to integrate numerically.

\subsection{1D-1V Vlasov Equation with Variance-based Greedy Sampling}
\label{sec:3vlasov}
\noindent In this last example, we introduce a GPLaSDI example to showcase the variance based active learning strategy. We consider the simplified 1D--1V Vlasov--Poisson equation, for which implementing the residual-based sampling strategy is cumbersome:
\begin{equation}
\label{vlasov_eqn}
\begin{cases}
    \displaystyle\frac{\partial f}{\partial t}
    +\frac{\partial }{\partial x} \left(v f\right) 
    + \frac{\partial}{\partial v}\left(\frac{d \phi}{dx} f\right) 
    = 0
    \hspace{0.3in}
    (t,x,v)\in[0,5]\times[0,2\pi]\times[-7,7]\\[15pt]
    \displaystyle\frac{d^2\phi}{dx^2} = \int_v f dv
\end{cases}
\end{equation}
$f$ is the plasma distribution function, dependant on a spatial coordinate $x$ and a velocity coordinate $v$, and $\phi$ is the electrostatic potential. This model describes collisionless electrostatic plasma dynamics within a 1-dimensional space, and is representative of more complex plasma behaviors occuring in nuclear fusion reactors. Due to the velocity variable, this equation can be seen (and solved) as a 2D-PDE. The initial condition is a two-stream instability problem described by the following expression:
\begin{equation}
    f(t=0,x,v)=\frac{4}{\pi T}\bigg[1+\frac{1}{10}\cos(k\pi x)\bigg]\bigg[\exp\bigg(-\frac{(v-2)^2}{2T}\bigg)+\exp\bigg(-\frac{(v+2)^2}{2T}\bigg)\bigg]
\end{equation}
We consider two simulation parameters, $T\in[0.9, 1.1]$ and $k\in[1.0,1.2]$ ($\pmb{\mu}=\{T,k\}$). The parameter space is discretized over a $21\times21$ grid $\mathcal{D}^h$, with a step size of $\Delta T=\Delta k=0.01$. The initial training data contains the four corners of the parameter space ($N_\mu=4$). To generate the FOM data, we employ \texttt{HyPar} \cite{HyPar}, a conservative finite difference PDE code with a WENO spatial discretization~\cite{jiangshu} and the classical fourth--order Runge--Kutta time integration scheme ($\Delta t=5\cdot10^{-3}$). The auto-encoder employs fully connected layers with softplus activation functions, and $N_z=5$ latent space variables. The SINDy library is restricted to linear and constant terms, which we have found to be sufficient for modeling the latent space dynamics (in this example, WLaSDI is not used). GPLaSDI is trained with $N_{epoch}=6.5\cdot10^5$ epochs, and a FOM data sampling rate every $N_{up}=5\cdot10^4$ epochs (resulting in adding 12 data points during training, for a total of 16 training points). More details on the models hyperparameters can be found in \cite{BONNEVILLE2024116535}.

\begin{figure}[!h]
\hspace{-1.cm}
    \includegraphics[width=1.1\textwidth]{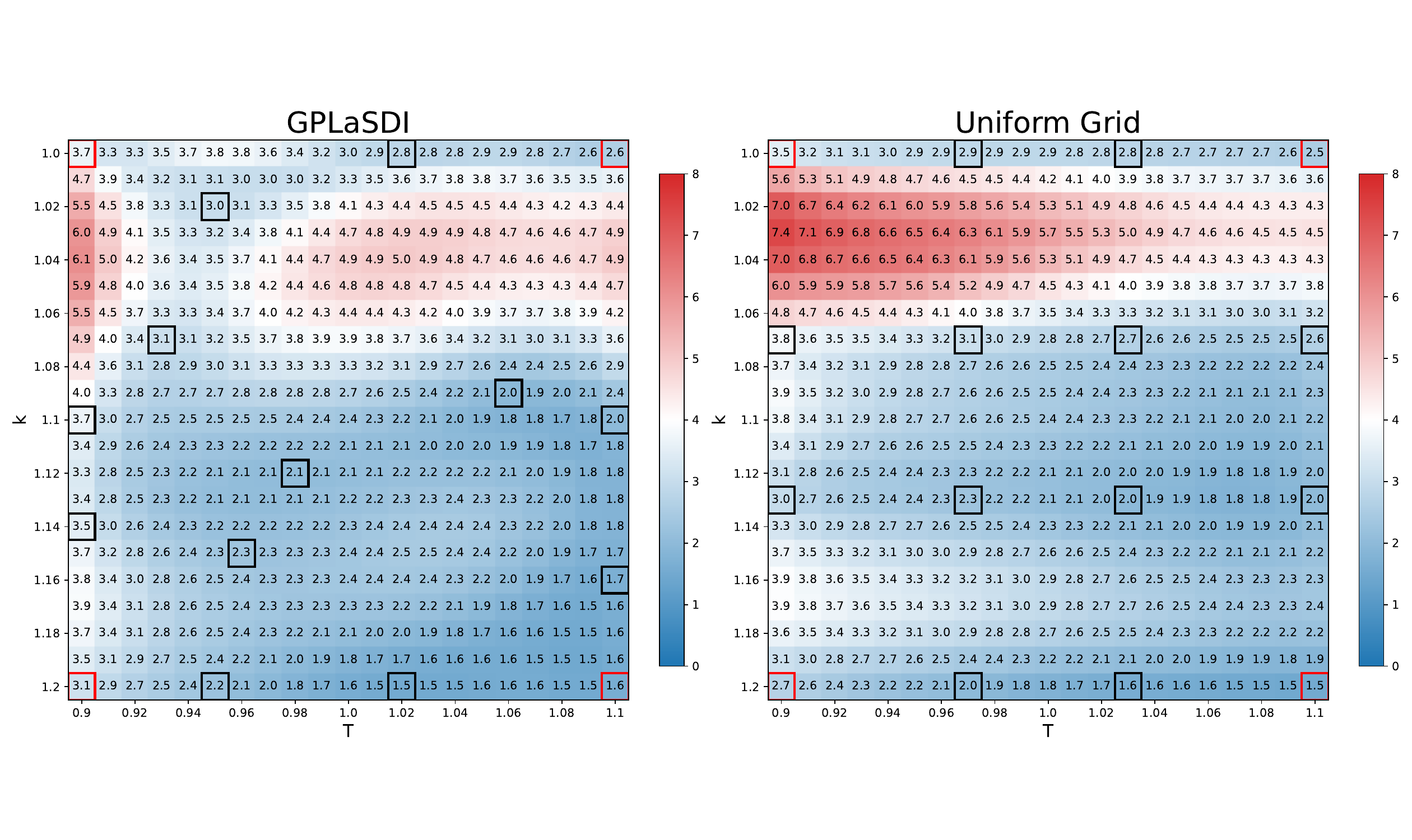}
    \caption{1D1V Vlasov Equation -- Maximum relative error ($\%$) using GPLaSDI and a uniform training grid (non--greedy).}
    \label{vlasov_max_error_mean}
\end{figure}

For baseline comparison, we also trained the same model but with pre-selected training parameters. The 16 training data points are associated with parameters located along a uniform $4\times4$ grid. All the training hyperparameters are kept identical and the baseline model also employs GPs for interpolating the latent space ODE coefficients. Figure \ref{vlasov_max_error_mean} presents the maximum relative error for each point in the parameter space obtained using GPLaSDI and the baseline model. With GPLaSDI, the worst maximum relative error is $6.1\%$, and in most regions of the parameter space, the error remains within the range of $1.5-3.5\%$. The highest errors are concentrated towards smaller values of $k$ (typically $k<1.07$). Compared to uniform sampling, GPLaSDI outperforms the baseline model, which achieves a maximum relative error of $7.4\%$. It is clear that the variance-based sampling reduces the error faster by selecting data where it is needed the most.

\begin{figure}[!h]
\hspace{-1.5cm}
        \includegraphics[width=1.25\textwidth]{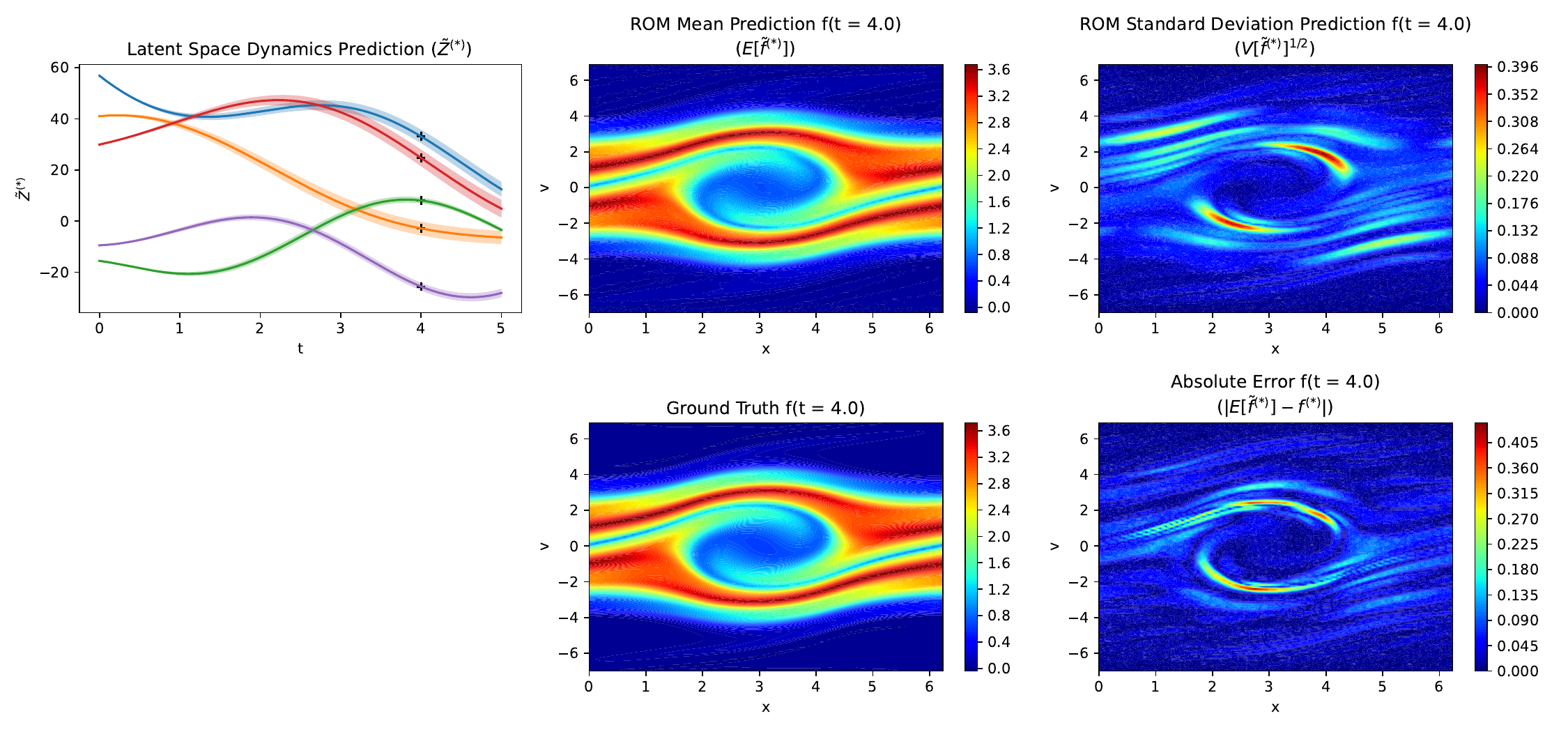}
    \caption{1D1V Vlasov Equation -- Prediction for $\pmb{\mu}^{(*)}=\{0.9,1.04\}$ at $t=4.0$. The figure illustrates the predicted latent space dynamics $\mathbb{E}[\mathbf{\tilde{Z}}^{(*)}]$ with a $95\%$ confidence interval, the ROM mean prediction $\mathbb{E}[\tilde{f}^{(*)}]$ and standard deviation $\mathbb{V}[\tilde{f}^{(*)}]^{1/2}$, the ground truth, and the absolute error.}
    \label{vlasov_prediction}
\end{figure}

Figure \ref{vlasov_prediction} displays the latent space dynamics, including the predicted and ground truth values of $f$, the absolute error, and the predictive standard deviation. The results correspond to the least favorable case ($\pmb{\mu}^{(*)}=\{0.9,1.04\}$) at $t=4$. The standard deviation of the reduced-order model (ROM) exhibits qualitative similarity to the absolute error, and the error generally falls within about 1-standard-deviation. This shows that GPLaSDI is able to output meaningful confidence intervals with well quantified uncertainty. 

In $20$ separate test runs, the FOM requires an average wall clock run--time of $22.5$ seconds when utilizing four cores, and $57.9$ seconds when using a single core. In contrast, the ROM model achieves an average run--time of $1.18\cdot10^{-2}$ seconds, resulting in a remarkable average speed-up of $4906\times$ ($1906\times$ when compared to the parallel FOM). Note that this speed-up is obtained using the mean prediction only, and thus does not permits uncertainty quantification (i.e. rather than sampling multiple times $\mathbf{\Xi}^{(d)}\sim\mathcal{N}(\mathbf{\Xi}^{(*)}|m^{(*)},s^{(*)2})$, we take $\mathbf{\Xi}^{(*)}=m^{(*)}$).

    
\section{Conclusion}
In this chapter, we have summarized the key building blocks that can be used and interchanged to develop LaSDI-based ROMs. By training an auto-encoder over high-fidelity data, we can effectively compress complex physics described by a PDE into simpler latent dynamics described by coupled ODEs. The latent space dynamics can be identified during the auto-encoder training using SINDy, Weak-SINDy, or the GENERIC framework. Interpolation can be applied to exploit the local latent space dynamics of training parameters for the prediction of testing parameters. Finally, active learning can be employed through either the residual-based error (gLaSDI for intrusive ROM) or the prediction uncertainty (GPLaSDI for non-intrusive ROM). LaSDI algorithms achieve remarkable performance, both in terms of accuracy (typically less than a $5$-$10\%$ maximum relative error) and efficiency (up to a few thousand times speed-up). The LaSDI framework can be applied to a wide variety of non-equilibrium physical problems, as we demonstrated with the Burgers equation, the non-linear heat equation, and the Vlasov equation. Recent development and extension have enabled LaSDI to be robust against noisy data (WLaSDI), to rigorously respect the laws of thermodynamics (tLaSDI), efficiently sample training data (gLaSDI, GPLaSDI), and provide meaningful prediction confidence intervals (GPLaSDI).

\begin{credits}
\subsubsection{\ackname} 
This work received support from the U.S. Department of Energy, Office of Science, Office of Advanced Scientific Computing Research, as part of the CHaRMNET Mathematical Multifaceted Integrated Capability Center (MMICC) program, under Award Number DE-SC0023164 to YC at Lawrence Livermore National Laboratory, and under Award Number DE-SC0023346 to DMB at the University of Colorado Boulder. The LLNL Lab Directed Research and Development Program is acknowledged for funding support of this research under Project No. 21-SI-006. 
JP was supported by a KIAS Individual Grant (AP095601) via the Center for AI and Natural Sciences at Korea Institute for Advanced Study.
YS was partially supported for this work by the NRF grant funded by the Ministry of Science and ICT of Korea (RS-2023-00219980).
Livermore National Laboratory is operated by Lawrence Livermore National Security, LLC, for the U.S. Department of Energy, National Nuclear Security Administration under Contract DE-AC52-07NA27344. IM release number: LLNL-BOOK-861682. 

\end{credits}
\bibliographystyle{splncs04}
\bibliography{references}

\begin{thebibliography}{10}
\providecommand{\url}[1]{\texttt{#1}}
\providecommand{\urlprefix}{URL }
\providecommand{\doi}[1]{https://doi.org/#1}

\bibitem{HyPar}
{HyPar} {R}epository, {\tt https://bitbucket.org/deboghosh/hypar}

\bibitem{Journal}
Review of digital twin about concepts, technologies, and industrial applications. Journal of Manufacturing Systems  \textbf{58},  346--361 (2020). \doi{10.1016/J.JMSY.2020.06.017}

\bibitem{mfem}
Anderson, R., Andrej, J., Barker, A., Bramwell, J., Camier, J.S., Cerveny, J., Dobrev, V., Dudouit, Y., Fisher, A., Kolev, T., Pazner, W., Stowell, M., Tomov, V., Akkerman, I., Dahm, J., Medina, D., Zampini, S.: {MFEM}: A modular finite element methods library. Computers \& Mathematics with Applications  \textbf{81},  42--74 (2021). \doi{10.1016/j.camwa.2020.06.009}

\bibitem{https://doi.org/10.48550/arxiv.2106.09658}
Bai, Z., Peng, L.: Non-intrusive nonlinear model reduction via machine learning approximations to low-dimensional operators (2021). \doi{10.48550/ARXIV.2106.09658}, \url{https://arxiv.org/abs/2106.09658}

\bibitem{doi:10.1146/annurev.fl.25.010193.002543}
Berkooz, G., Holmes, P., Lumley, J.L.: The proper orthogonal decomposition in the analysis of turbulent flows. Annual Review of Fluid Mechanics  \textbf{25}(1),  539--575 (1993). \doi{10.1146/annurev.fl.25.010193.002543}, \url{https://doi.org/10.1146/annurev.fl.25.010193.002543}

\bibitem{10754/656260}
Biegler, L., Biros, G., Ghattas, O., Heinkenschloss, M., Keyes, D., Mallick, B., Marzouk, Y., Tenorio, L., van Bloemen~Waanders, B., Willcox, K.: Large-scale inverse problems and quantification of uncertainty (2010). \doi{10.1002/9780470685853}, \url{http://hdl.handle.net/10754/656260}

\bibitem{bonneville2023datadriven}
Bonneville, C., Choi, Y., Ghosh, D., Belof, J.L.: Data-driven autoencoder numerical solver with uncertainty quantification for fast physical simulations (2023)

\bibitem{BONNEVILLE2024116535}
Bonneville, C., Choi, Y., Ghosh, D., Belof, J.L.: Gplasdi: Gaussian process-based interpretable latent space dynamics identification through deep autoencoder. Computer Methods in Applied Mechanics and Engineering  \textbf{418},  116535 (2024). \doi{https://doi.org/10.1016/j.cma.2023.116535}, \url{https://www.sciencedirect.com/science/article/pii/S004578252300659X}

\bibitem{BONNEVILLE2022100115}
Bonneville, C., Earls, C.: Bayesian deep learning for partial differential equation parameter discovery with sparse and noisy data. Journal of Computational Physics: X  \textbf{16},  100115 (2022). \doi{https://doi.org/10.1016/j.jcpx.2022.100115}, \url{https://www.sciencedirect.com/science/article/pii/S2590055222000117}

\bibitem{wendy}
Bortz, D.M., Messenger, D.A., Dukic, V.: Direct {{Estimation}} of {{Parameters}} in {{ODE Models Using WENDy}}: {{Weak-form Estimation}} of {{Nonlinear Dynamics}}. Bull. Math. Biol.  \textbf{85}(110) (2023). \doi{DOI:10.1007/S11538-023-01208-6}

\bibitem{doi:10.1073/pnas.1517384113}
Brunton, S.L., Proctor, J.L., Kutz, J.N.: Discovering governing equations from data by sparse identification of nonlinear dynamical systems. Proceedings of the National Academy of Sciences  \textbf{113}(15),  3932--3937 (2016). \doi{10.1073/pnas.1517384113}, \url{https://www.pnas.org/doi/abs/10.1073/pnas.1517384113}

\bibitem{article}
Calder, M., Craig, C., Culley, D., Cani, R., Donnelly, C., Douglas, R., Edmonds, B., Gascoigne, J., Gilbert, N., Hargrove, C., Hinds, D., Lane, D., Mitchell, D., Pavey, G., Robertson, D., Rosewell, B., Sherwin, S., Walport, M., Wilson, A.: Computational modelling for decision-making: Where, why, what, who and how. Royal Society Open Science  \textbf{5},  172096 (06 2018). \doi{10.1098/rsos.172096}

\bibitem{doi:10.1073/pnas.1906995116}
Champion, K., Lusch, B., Kutz, J.N., Brunton, S.L.: Data-driven discovery of coordinates and governing equations. Proceedings of the National Academy of Sciences  \textbf{116}(45),  22445--22451 (2019). \doi{10.1073/pnas.1906995116}, \url{https://www.pnas.org/doi/abs/10.1073/pnas.1906995116}

\bibitem{Chen_2021}
Chen, Z., Liu, Y., Sun, H.: Physics-informed learning of governing equations from scarce data. Nature Communications  \textbf{12}(1) (oct 2021). \doi{10.1038/s41467-021-26434-1}, \url{https://doi.org/10.1038/s41467-021-26434-1}

\bibitem{unknown}
Cheng, K., Zimmermann, R.: Sliced gradient-enhanced kriging for high-dimensional function approximation and aerodynamic modeling (04 2022)

\bibitem{https://doi.org/10.48550/arxiv.2201.07335}
Cheung, S.W., Choi, Y., Copeland, D.M., Huynh, K.: Local lagrangian reduced-order modeling for rayleigh-taylor instability by solution manifold decomposition (2022). \doi{10.48550/ARXIV.2201.07335}, \url{https://arxiv.org/abs/2201.07335}

\bibitem{doi:10.1137/19M1242963}
Choi, Y., Coombs, D., Anderson, R.: Sns: A solution-based nonlinear subspace method for time-dependent model order reduction. SIAM Journal on Scientific Computing  \textbf{42}(2),  A1116--A1146 (2020). \doi{10.1137/19M1242963}, \url{https://doi.org/10.1137/19M1242963}

\bibitem{oc}
Choi, Y., Farhat, C., Murray, W., Saunders, M.: A practical factorization of a schur complement for pde-constrained distributed optimal control (2013). \doi{10.48550/ARXIV.1312.5653}, \url{https://arxiv.org/abs/1312.5653}

\bibitem{Copeland_2022}
Copeland, D.M., Cheung, S.W., Huynh, K., Choi, Y.: Reduced order models for lagrangian hydrodynamics. Computer Methods in Applied Mechanics and Engineering  \textbf{388},  114259 (jan 2022). \doi{10.1016/j.cma.2021.114259}, \url{https://doi.org/10.1016/j.cma.2021.114259}

\bibitem{cummings_mason_morton_mcdaniel_2015}
Cummings, R.M., Mason, W.H., Morton, S.A., McDaniel, D.R.: Applied Computational Aerodynamics: A Modern Engineering Approach. Cambridge Aerospace Series, Cambridge University Press (2015). \doi{10.1017/CBO9781107284166}

\bibitem{NIPS1992_cdc0d6e6}
DeMers, D., Cottrell, G.: Non-linear dimensionality reduction. In: Hanson, S., Cowan, J., Giles, C. (eds.) Advances in Neural Information Processing Systems. vol.~5. Morgan-Kaufmann (1992), \url{https://proceedings.neurips.cc/paper/1992/file/cdc0d6e63aa8e41c89689f54970bb35f-Paper.pdf}

\bibitem{diaz2023fast}
Diaz, A.N., Choi, Y., Heinkenschloss, M.: A fast and accurate domain-decomposition nonlinear manifold reduced order model (2023)

\bibitem{6db924dfeff44d159ab577c1aefed6ef}
Diston, D.: Computational Modelling and Simulation of Aircraft and the Environment: Platform Kinematics and Synthetic Environment, Aerospace Series, vol.~1. John Wiley \& Sons Ltd, United Kingdom, 1 edn. (Apr 2009). \doi{10.1002/9780470744130}

\bibitem{https://doi.org/10.1002/2016RS005998}
Fountoulakis, V., Earls, C.: Duct heights inferred from radar sea clutter using proper orthogonal bases. Radio Science  \textbf{51}(10),  1614--1626 (2016). \doi{https://doi.org/10.1002/2016RS005998}, \url{https://agupubs.onlinelibrary.wiley.com/doi/abs/10.1002/2016RS005998}

\bibitem{lasdi}
Fries, W.D., He, X., Choi, Y.: {LaSDI}: Parametric latent space dynamics identification. Computer Methods in Applied Mechanics and Engineering  \textbf{399},  115436 (sep 2022). \doi{10.1016/j.cma.2022.115436}, \url{https://doi.org/10.1016/j.cma.2022.115436}

\bibitem{https://doi.org/10.1002/nme.2746}
Galbally, D., Fidkowski, K., Willcox, K., Ghattas, O.: Non-linear model reduction for uncertainty quantification in large-scale inverse problems. International Journal for Numerical Methods in Engineering  \textbf{81}(12),  1581--1608 (2010). \doi{https://doi.org/10.1002/nme.2746}, \url{https://onlinelibrary.wiley.com/doi/abs/10.1002/nme.2746}

\bibitem{https://doi.org/10.48550/arxiv.2211.10575}
Gao, L.M., Kutz, J.N.: Bayesian autoencoders for data-driven discovery of coordinates, governing equations and fundamental constants (2022). \doi{10.48550/ARXIV.2211.10575}, \url{https://arxiv.org/abs/2211.10575}

\bibitem{Goodfellow-et-al-2016}
Goodfellow, I., Bengio, Y., Courville, A.: Deep Learning. MIT Press (2016), \url{http://www.deeplearningbook.org}

\bibitem{grmela1997dynamics}
Grmela, M., {\"O}ttinger, H.C.: Dynamics and thermodynamics of complex fluids. i. development of a general formalism. Physical Review E  \textbf{56}(6), ~6620 (1997)

\bibitem{ha2017hypernetworks}
Ha, D., Dai, A.M., Le, Q.V.: Hypernetworks. In: International Conference on Learning Representations (2017), \url{https://openreview.net/forum?id=rkpACe1lx}

\bibitem{glasdi}
He, X., Choi, Y., Fries, W.D., Belof, J., Chen, J.S.: {gLaSDI}: Parametric physics-informed greedy latent space dynamics identification  \textbf{489},  112267 (2023). \doi{10.1016/j.jcp.2023.112267}, \url{https://www.sciencedirect.com/science/article/abs/pii/S0021999123003625}

\bibitem{he2022certified}
He, X., Choi, Y., Fries, W.D., Belof, J.L., Chen, J.S.: Certified data-driven physics-informed greedy auto-encoder simulator (2022)

\bibitem{hernandez2021structure}
Hernandez, Q., Badias, A., Gonzalez, D., Chinesta, F., Cueto, E.: Structure-preserving neural networks. Journal of Computational Physics  \textbf{426},  109950 (2021)

\bibitem{doi:10.1126/science.1127647}
Hinton, G.E., Salakhutdinov, R.R.: Reducing the dimensionality of data with neural networks. Science  \textbf{313}(5786),  504--507 (2006). \doi{10.1126/science.1127647}, \url{https://www.science.org/doi/abs/10.1126/science.1127647}

\bibitem{doi:10.1098/rsos.211823}
Hirsh, S.M., Barajas-Solano, D.A., Kutz, J.N.: Sparsifying priors for bayesian uncertainty quantification in model discovery. Royal Society Open Science  \textbf{9}(2),  211823 (2022). \doi{10.1098/rsos.211823}, \url{https://royalsocietypublishing.org/doi/abs/10.1098/rsos.211823}

\bibitem{https://doi.org/10.1002/num.21835}
Iliescu, T., Wang, Z.: Variational multiscale proper orthogonal decomposition: Navier-stokes equations. Numerical Methods for Partial Differential Equations  \textbf{30}(2),  641--663 (2014). \doi{https://doi.org/10.1002/num.21835}, \url{https://onlinelibrary.wiley.com/doi/abs/10.1002/num.21835}

\bibitem{jiangshu}
Jiang, G.S., Shu, C.W.: Efficient implementation of weighted {ENO} schemes. Journal of Computational Physics  \textbf{126}(1),  202--228 (1996). \doi{10.1006/jcph.1996.0130}

\bibitem{JONES202036}
Jones, D., Snider, C., Nassehi, A., Yon, J., Hicks, B.: Characterising the digital twin: A systematic literature review. CIRP Journal of Manufacturing Science and Technology  \textbf{29},  36--52 (2020). \doi{https://doi.org/10.1016/j.cirpj.2020.02.002}, \url{https://www.sciencedirect.com/science/article/pii/S1755581720300110}

\bibitem{ident}
Kang, S.H., Liao, W., Liu, Y.: {{IDENT}}: {{Identifying Differential Equations}} with {{Numerical Time Evolution}}. J Sci Comput  \textbf{87}(1), ~1 (Apr 2021). \doi{10.1007/s10915-020-01404-9}

\bibitem{https://doi.org/10.48550/arxiv.2011.07727}
Kim, Y., Choi, Y., Widemann, D., Zohdi, T.: Efficient nonlinear manifold reduced order model (2020). \doi{10.48550/ARXIV.2011.07727}, \url{https://arxiv.org/abs/2011.07727}

\bibitem{https://doi.org/10.48550/arxiv.2009.11990}
Kim, Y., Choi, Y., Widemann, D., Zohdi, T.: A fast and accurate physics-informed neural network reduced order model with shallow masked autoencoder (2020). \doi{10.48550/ARXIV.2009.11990}, \url{https://arxiv.org/abs/2009.11990}

\bibitem{KimChoiWidemannEtAl2022JournalofComputationalPhysics}
Kim, Y., Choi, Y., Widemann, D., Zohdi, T.: A fast and accurate physics-informed neural network reduced order model with shallow masked autoencoder. Journal of Computational Physics  \textbf{451},  110841 (Feb 2022). \doi{10.1016/j.jcp.2021.110841}

\bibitem{math9141690}
Kim, Y., Wang, K., Choi, Y.: Efficient space–time reduced order model for linear dynamical systems in python using less than 120 lines of code. Mathematics  \textbf{9}(14) (2021). \doi{10.3390/math9141690}, \url{https://www.mdpi.com/2227-7390/9/14/1690}

\bibitem{car1}
Kurec, K., Remer, M., Broniszewski, J., Bibik, P., Tudruj, S., Piechna, J.: Advanced modeling and simulation of vehicle active aerodynamic safety. Journal of Advanced Transportation  \textbf{2019},  1--17 (02 2019). \doi{10.1155/2019/7308590}

\bibitem{kutz_2017}
Kutz, J.N.: Deep learning in fluid dynamics. Journal of Fluid Mechanics  \textbf{814},  1–4 (2017). \doi{10.1017/jfm.2016.803}

\bibitem{https://doi.org/10.48550/arxiv.2203.16494}
Lauzon, J.T., Cheung, S.W., Shin, Y., Choi, Y., Copeland, D.M., Huynh, K.: S-opt: A points selection algorithm for hyper-reduction in reduced order models (2022). \doi{10.48550/ARXIV.2203.16494}, \url{https://arxiv.org/abs/2203.16494}

\bibitem{LEE2020108973}
Lee, K., Carlberg, K.T.: Model reduction of dynamical systems on nonlinear manifolds using deep convolutional autoencoders. Journal of Computational Physics  \textbf{404},  108973 (2020). \doi{https://doi.org/10.1016/j.jcp.2019.108973}, \url{https://www.sciencedirect.com/science/article/pii/S0021999119306783}

\bibitem{lee2021machine}
Lee, K., Trask, N., Stinis, P.: Machine learning structure preserving brackets for forecasting irreversible processes. Advances in Neural Information Processing Systems  \textbf{34},  5696--5707 (2021)

\bibitem{Marjavaara2006CFDDO}
Marjavaara, D.: Cfd driven optimization of hydraulic turbine draft tubes using surrogate models (2006)

\bibitem{mcbane2022stress}
McBane, S., Choi, Y., Willcox, K.: Stress-constrained topology optimization of lattice-like structures using component-wise reduced order models. Computer Methods in Applied Mechanics and Engineering  \textbf{400},  115525 (2022)

\bibitem{MCLAUGHLIN20162407}
McLaughlin, B., Peterson, J., Ye, M.: Stabilized reduced order models for the advection–diffusion–reaction equation using operator splitting. Computers \& Mathematics with Applications  \textbf{71}(11),  2407--2420 (2016). \doi{https://doi.org/10.1016/j.camwa.2016.01.032}, \url{https://www.sciencedirect.com/science/article/pii/S0898122116300281}, proceedings of the conference on Advances in Scientific Computing and Applied Mathematics. A special issue in honor of Max Gunzburger’s 70th birthday

\bibitem{wsindy2}
Messenger, D.A., Bortz, D.M.: Weak {{SINDy For Partial Differential Equations}}. J. Comput. Phys.  \textbf{443},  110525 (Oct 2021). \doi{10.1016/j.jcp.2021.110525}

\bibitem{wsindy1}
Messenger, D.A., Bortz, D.M.: Weak {{SINDy}}: {{Galerkin-Based Data-Driven Model Selection}}. Multiscale Model. Simul.  \textbf{19}(3),  1474--1497 (2021). \doi{10.1137/20M1343166}

\bibitem{9043275}
Muhammad, A., Shanono, I.H.: Simulation of a car crash using ansys. In: 2019 15th International Conference on Electronics, Computer and Computation (ICECCO). pp.~1--5 (2019). \doi{10.1109/ICECCO48375.2019.9043275}

\bibitem{ottinger2005beyond}
{\"O}ttinger, H.C.: Beyond equilibrium thermodynamics. John Wiley \& Sons, Hoboken, NJ (2005)

\bibitem{park2024tlasdi}
Park, J.S.R., Cheung, S.W., Choi, Y., Shin, Y.: t{L}a{SDI}: Thermodynamics-informed latent space dynamics identification (2024), \url{https://arxiv.org/abs/2403.05848}

\bibitem{Peterson_b1998}
Peterson, A.F., Ray, S.L., Mittra, R.:

\bibitem{alma991043311449403276}
Raczynski, S.: Modeling and simulation : the computer science of illusion / Stanislaw Raczynski. RSP Series in Computer Simulation and Modeling, John Wiley \& Sons, Ltd, Hertfordshire, England ; (2006 - 2006)

\bibitem{books/lib/RasmussenW06}
Rasmussen, C.E., Williams, C.K.I.: Gaussian processes for machine learning. Adaptive computation and machine learning, MIT Press (2006)

\bibitem{rbm}
Rozza, G., Huynh, D., Patera, A.: Reduced basis approximation and a posteriori error estimation for affinely parametrized elliptic coercive partial differential equations. Archives of Computational Methods in Engineering  \textbf{15},  1--47 (09 2007). \doi{10.1007/BF03024948}

\bibitem{doi:10.1126/sciadv.1602614}
Rudy, S.H., Brunton, S.L., Proctor, J.L., Kutz, J.N.: Data-driven discovery of partial differential equations. Science Advances  \textbf{3}(4),  e1602614 (2017). \doi{10.1126/sciadv.1602614}, \url{https://www.science.org/doi/abs/10.1126/sciadv.1602614}

\bibitem{Safonov1988ASM}
Safonov, M.G., Chiang, R.Y.: A schur method for balanced model reduction. 1988 American Control Conference pp. 1036--1040 (1988)

\bibitem{russel}
Schwartz, R.: Biological Modeling and Simulation. MIT Press (2008)

\bibitem{Smith2013UncertaintyQ}
Smith, R.C.: Uncertainty quantification - theory, implementation, and applications. In: Computational science and engineering (2013)

\bibitem{Stabile_2018}
Stabile, G., Rozza, G.: Finite volume {POD}-galerkin stabilised reduced order methods for the parametrised incompressible navier{\textendash}stokes equations. Computers \& Fluids  \textbf{173},  273--284 (sep 2018). \doi{10.1016/j.compfluid.2018.01.035}, \url{https://doi.org/10.1016/j.compfluid.2018.01.035}

\bibitem{STEPHANY2022360}
Stephany, R., Earls, C.: Pde-read: Human-readable partial differential equation discovery using deep learning. Neural Networks  \textbf{154},  360--382 (2022). \doi{https://doi.org/10.1016/j.neunet.2022.07.008}, \url{https://www.sciencedirect.com/science/article/pii/S0893608022002660}

\bibitem{Sternfels_2013}
Sternfels, R., Earls, C.J.: Reduced-order model tracking and interpolation to solve pde-based bayesian inverse problems. Inverse Problems  \textbf{29}(7),  075014 (jun 2013). \doi{10.1088/0266-5611/29/7/075014}, \url{https://dx.doi.org/10.1088/0266-5611/29/7/075014}

\bibitem{weakident}
Tang, M., Liao, W., Kuske, R., Kang, S.H.: {{WeakIdent}}: {{Weak}} formulation for {{Identifying Differential Equation}} using {{Narrow-fit}} and {{Trimming}}. J. Comput. Phys. p. 112069 (Mar 2023). \doi{10.1016/j.jcp.2023.112069}

\bibitem{osti_1420279}
Tapia, G., Khairallah, S.A., Matthews, M.J., King, W.E., Elwany, A.: Gaussian process-based surrogate modeling framework for process planning in laser powder-bed fusion additive manufacturing of 316l stainless steel. International Journal of Advanced Manufacturing Technology  \textbf{94}(9-12) (9 2017). \doi{10.1007/s00170-017-1045-z}

\bibitem{thijssen_2007}
Thijssen, J.: Computational Physics. Cambridge University Press, 2 edn. (2007). \doi{10.1017/CBO9781139171397}

\bibitem{rylander}
Thomas~Rylander, Par~Ingelström, A.B.: Computational Electromagnetics. Springer (2013)

\bibitem{wlasdi}
Tran, A., He, X., Messenger, D.A., Choi, Y., Bortz, D.M.: Weak-{{Form Latent Space Dynamics Identification}}. Comput. Methods Appl. Mech. Eng. (accepted) (2024)

\bibitem{do1}
Wang, S., Sturler, E.d., Paulino, G.H.: Large-scale topology optimization using preconditioned krylov subspace methods with recycling. International Journal for Numerical Methods in Engineering  \textbf{69}(12),  2441--2468 (2007). \doi{https://doi.org/10.1002/nme.1798}, \url{https://onlinelibrary.wiley.com/doi/abs/10.1002/nme.1798}

\bibitem{do2}
White, D., Choi, Y., Kudo, J.: A dual mesh method with adaptivity for stress-constrained topology optimization. Structural and Multidisciplinary Optimization  \textbf{61} (02 2020). \doi{10.1007/s00158-019-02393-6}

\bibitem{sep-simulations-science}
Winsberg, E.: {Computer Simulations in Science}. In: Zalta, E.N., Nodelman, U. (eds.) The {Stanford} Encyclopedia of Philosophy. Metaphysics Research Lab, Stanford University, {W}inter 2022 edn. (2022)

\bibitem{zhang2022gfinns}
Zhang, Z., Shin, Y., Karniadakis, G.E.: {GFINN}s: Generic formalism informed neural networks for deterministic and stochastic dynamical systems. Philosophical Transactions of the Royal Society A  \textbf{380}(2229),  20210207 (2022)

\end{thebibliography}

%
%
%

%




\end{document}